# SPATIOTEMPORAL SENSITIVITY AND VISUAL ATTENTION FOR EFFICIENT RENDERING OF DYNAMIC ENVIRONMENTS

A Thesis

Presented to the Faculty of the Graduate School

of Cornell University

in Partial Fulfillment of the Requirements for the Degree of

Master of Science

by

Yang Li Hector Yee

August 2000

# ABSTRACT

We present a method to accelerate global illumination computation in dynamic environments by taking advantage of limitations of the human visual system. A model of visual attention is used to locate regions of interest in a scene and to modulate spatiotemporal sensitivity. The method is applied in the form of a spatiotemporal error tolerance map. Perceptual acceleration combined with good sampling protocols provide a global illumination solution feasible for use in animation. Results indicate an order of magnitude improvement in computational speed. The method is adaptable and can also be used in image-based rendering, geometry level of detail selection, realistic image synthesis, video telephony and video compression.

# BIOGRAPHICAL SKETCH

The author was born December 15, 1974 in the Republic of Singapore. He studied computer science with a minor in business at Cornell University. In 1998, he graduated with a Bachelor of Science degree and was accepted as a graduate student into the Program of Computer Graphics in the Field of Architecture at Cornell University. He graduated from Cornell University in August, 2000 with a Master of Science degree in computer graphics.

To my family.

# ACKNOWLEDGEMENTS

I would like to thank Donald Greenberg for giving me this wonderful opportunity of working in the Cornell Program of Computer Graphics. His vision and support was invaluable in creating a productive and enjoyable atmosphere for constructive work at the lab.

Many thanks go to Sumant Pattanaik for being my mentor and guide in the realm of perception and vision science. His help was invaluable and made my thesis possible.

I would like to thank Phil Dutre for listening to my endless questions and wacky ideas for global illumination and for painstakingly teaching me the rendering equation a zillion times. I found his global illumination compendium a very useful reference in checking the veracity of my global illumination code. He also gave me the idea of calling the spatiotemporal sensitivity map the Aleph Map rather than some long and cumbersome A.C.R.O.N.Y.M. Thanks also to Bruce Walter and Jack Tumblin for putting up with the long discussions in their room and for joining in with the technical conversations. Speaking of long discussions, thanks go to Stephen Westin for regaling me with tales of how the old-timers derived Bi-directional Reflectance Distribution functions and to Ben Trumbore for providing me with comic relief afterwords. Jim Fewerda, Kavita 'KB' Bala, Reynald Dumont, Fabio 'Portabella' Pellacini, Sebestian 'Spiff' Fernandez, Mashesh 'Chapati' Ramasubramanian and Parag 'Chole' Tole were also invaluable resources in bouncing ideas off. Thanks go to David Hart who wrote some of the underlying code in my ren-derer, and to Greg Ward and Charles 'Chas' Erlich for helping me reconstruct the irradiance caching code from RADIANCE.

This work was facilitated by the aid of the staff in the Program of Computer Graphics. Thanks go to Jonathan Corson-Rikert, Linda Stephenson, Peggy Anderson, Martin Berggren and Mitch Collinsworth for making sure that the PCG machine is well oiled. Many thanks to Mary for coming in at 3 AM every morning to ensure that the office is clean.

The joy of the lab comes from Su-Anne 'Suzie' Fu, Corey Toler, Bryan Vandrovec, Sherwin 'Shervinator' Tam, Dan 'Raven-311' Gelb, Dan Kartch, Moreno Piccolotto, Michael Malone, Eric Wong, the 'Siamese twins' Steve Berman and Richard Levy, Randima 'Randy' Fernandez, Li Hongsong and the others I hang out with. Thanks to Eric Klatt for ensuring that I get my weekly exercise at the gym and Seth Kromholz for keeping the apartment tidy while I was writing my thesis. Thanks go to my mom, Janet Yee, for mailing me spices that I use to make Singaporean food and for the long distance phone support.

The animation sequences would not have been possible without the original artwork of Zehna Barros and Nordica Raapana. I also used free models from Viewpoint Datalabs and 3D cafe. Computation for this work was performed on workstations and compute clusters donated by Intel Corporation. Research was conducted in part using the resources of the Cornell Theory Center, which receives funding from Cornell University, New York State, federal agencies, and corporate partners.

This work was supported by the NSF Science and Technology Center for Computer Graphics and Scientific Visualization (ASC-8920219).

# TABLE OF CONTENTS

# LIST OF FIGURES

# CHAPTER 1
# Introduction

*ren-der* - *To give what is due or owed; To reduce, convert, or melt down fat by heating; To represent in a drawing or painting; - from dictionary.com*

In the quest for realism, computer graphics technology has evolved over the years from simple vector line drawings to the current state-of-the-art photorealistic images that are indistinguishable from photographs of the real world. The price of increased realism is the corresponding increase in computer time required to generate these images. Rendering a single realistic image frame can take many days even on the fastest computers of the new millennium (2000 AD).

One class of rendering algorithms used to generate realistic images are those that perform *global illumination*, the calculation of light transport in an environment [Gree97]. Figure 1.1 shows an example of an environment rendered with the global illumination technique. Many effects such as caustics, area light sources, soft shadows, anti-aliasing, motion blur and color bleeding are a natural result of the use of global illumination and do not require specific user intervention to construct. Global illumination generated images are physically accurate and are computed with high dynamic range. When measured data are used for the geometry and surface properties of objects in a scene, the image produced by a global illumination algorithm is practically indistinguishable from reality.

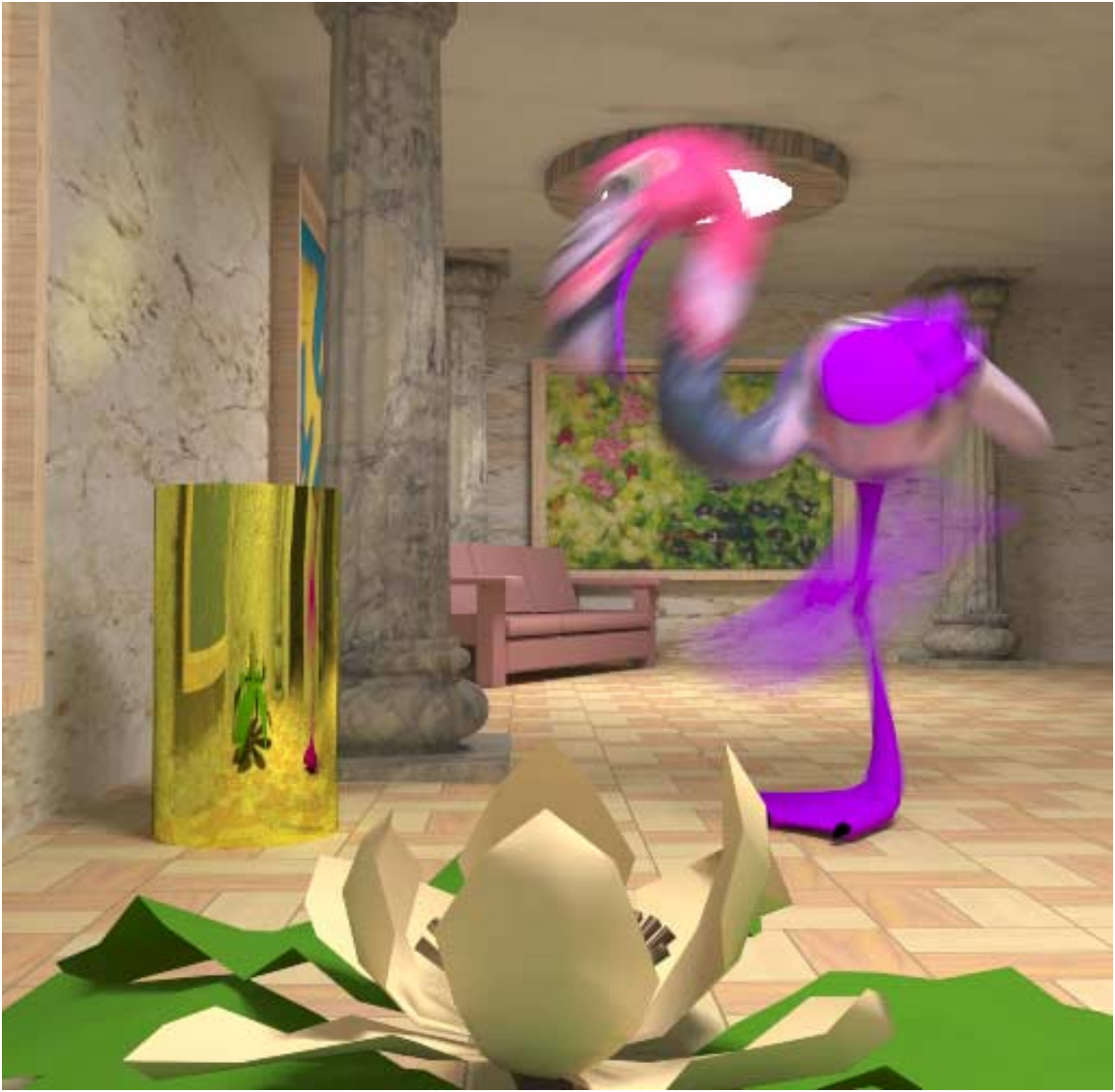

Figure 1.1: Global Illumination of a Dynamic Environment

Global illumination correctly simulates effects such as color bleeding (the green of the leaves on to the petals), motion blur (the beak and leg of the pink flamingo), caustics (the reflection of the light by the golden ash tray on the wall), soft shadows (the gradual change between the umbra and penumbra of the golden ash tray on the wall), anti-aliasing, and area light sources. The expensive operation of calculating a lighting solution for an environment benefits greatly from our new technique, which can be applied to animation as well as motion-blurred still images such as shown above.

Global illumination algorithms work by solving the rendering equation proposed by Kajiya [Kaji86]:

$$L_{\text{Out}} = L_{\text{E}} + \int\limits_{\Omega} L_{\text{In}} f_{\text{r}} \cos(\theta) d\omega_{\theta} \tag{1.1}$$

where $L_{\text{Out}}$ is the radiance leaving a surface, $L_{\text{E}}$ is the radiance emitted radiance by the surface, $L_{\text{In}}$ is the radiance of an incoming light ray arriving at the surface from light sources and other surfaces (e.g. reflector R), $f_{\text{r}}$ is the bi-directional reflection distribution function of the surface, $\theta$ is the angle between the surface normal and the incoming light ray, and $d\omega_{\theta}$ is the differential solid angle around the incoming light ray.

The equation essentially states that the light arriving at the eye is the sum of the light emitted from the surface, as well as the light reflected off the surface from light sources or other reflecting surfaces. The rendering equation is graphically depicted in Figure 1.2.

In Figure 1.2, $L_{\text{In}}$ is an example of a direct light source, such as a light bulb, while $L'_{\text{In}}$ is an example of an indirect light source, which can be light reflected from another reflecting surface such as a wall. To compute the global illumination of an environment, the rendering equation is recursively applied to reflecting surfaces like R as well as on all other surfaces and light sources in the environment. The light seen by the eye, $L_{\text{Out}}$, is simply the integral of the indirect and direct light sources modulated by the reflectance function of the surface over the hemisphere $\Omega$.

Let us consider the approximate expense of a fully-converged, global illumination solution in terms of year 2000 (Y2K) technology. A typical Y2K CRT monitor would have on the order of 1,000,000 pixels. Assuming a scene complexity of 100,000 polygons, a voxel-grid intersection algorithm takes

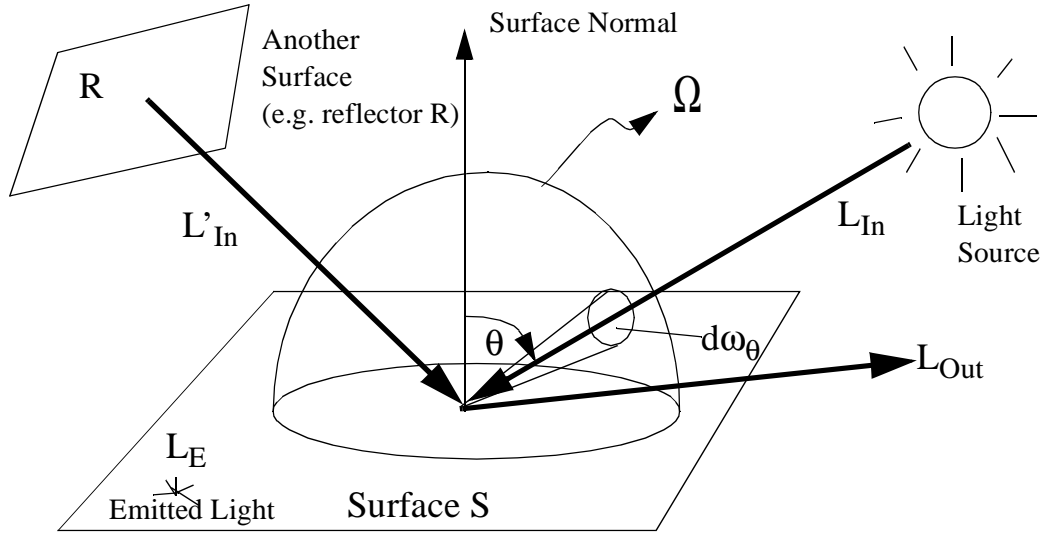

Figure 1.2: Graphical Depiction of the Rendering Equation

The lighting solution is being computed for surface S, with light coming from direct light sources or indirect light sources such as another surface, R.

about 100 cycles/sample to compute a single ray-triangle intersection. Since a voxel-grid intersection has logarithmic performance, and adding in the overhead of calculating shading, we estimate the cost for a single sample to be 1,000 cycles/sample. In order to get a good quality image, the global illumination algorithm would perform Monte Carlo integration on the rendering equation by shooting about 1,000 samples per pixel with an average path depth of 10 for a total of 10,000 samples per pixel. A typical Y2K processor runs at the speed of 1 GHz. Therefore a single image frame would take:

$$10^6 \text{ pixels} * 10^4 \text{ samples/pixel} * (10^3 \text{ cycles/sample}) / 1 \text{ GHz} = 10^4 \text{ s} \quad (1.2)$$

From the equation above, we estimate the time taken to compute a single converged image to be $10^4$ s or about 3 hours per frame. On a quad-processor compute node, it would take about 45 minutes to compute a fully converged image using Y2K ballpark figures for computer power and display image sizes. These are timing figures for a *single frame*. In dynamic environments, where moving objects and lights can potentially affect the lighting solution of every other object in the environment, a global illumination algorithm is forced to recompute the lighting solution anew for every frame in order to get a correct solution. Clearly, using a naive global illumination algorithm would be impractical for rendering dynamic environments, given the fact that the 1,000 frames for a 33 second video would take *125 days* to compute on a uniprocessor system! Even Moore's Law would not be useful for some time; therefore, we have to turn to alternative algorithms in order to practically apply global illumination algorithms to rendering dynamic environments.

There are a few ways to speed up the lighting computation of a dynamic environment. The brute force approach would be to throw more computer power at the problem by distributing the rendering calculation over a cluster of linked computers. This would buy us an order of magnitude improvement. Another order of magnitude speedup can be achieved by finding some way to exploit the spatiotemporal coherence inherent in an animation by using some kind of interpolation scheme. Lastly, our results have shown that intelligently applying a *perceptual oracle* to guide a global illumination algorithm would buy us another crucial order of magnitude in order to reduce the computation time of a 33 second video, by three orders of magnitude, from 125 days to a manageable 3 hours.

This thesis deals with the *perceptually-based rendering* of dynamic environments. The major assumption is that rendering is an expensive operation, and saving on computation costs in any way is substantially beneficial. One of the ways to save on computation costs is by the use of *perceptual error metrics*, which are error metrics based on computational models of the human visual system. Error metrics operate on two intermediate images of a global illumination solution in order to determine if the visual system is able to differentiate these two images. In this way, these perceptual metrics assist in rendering by informing a rendering algorithm when and where it can stop an iterative calculation prematurely because one may chose to stop solving when the differences are not perceptually noticeable. In doing so, perceptually-based renderers attempt to expend the least amount of work to obtain an image that is perceptually indistinguishable from a fully converged solution. A *perceptual oracle*, such as the one described in this thesis, is slightly different from a perceptual error metric. A perceptual oracle does not explicitly compute an error bound for a stopping condition by computing error differences, but instead provides perceptual information in advance of any global illumination computation in order to determine the most efficient way to render an image. The technique described in this paper is a perceptual oracle that assists rendering algorithms by producing a spatiotemporal error tolerance map (Aleph/ℵ Map) that can be used as a guide to optimize rendering. Figure 1.3 depicts an overview of the perceptual model used to generate the Aleph Map. It is called the Aleph Map because it is short for Application Adapted Attention Modulated Spatiotemporal Sensitivity Map.

Two psychophysical concepts are harnessed to generate the Aleph Map: spatiotemporal sensitivity and visual attention. The former tell us *how much error* we can tolerate and the latter, *where we look*. Sensitivity is important because it allows us to save on computation in areas where the eye is less sensitive and visual attention is important because it allows us to use sensitivity information wisely. Areas where attention is focused must be rendered more accurately than less important regions.

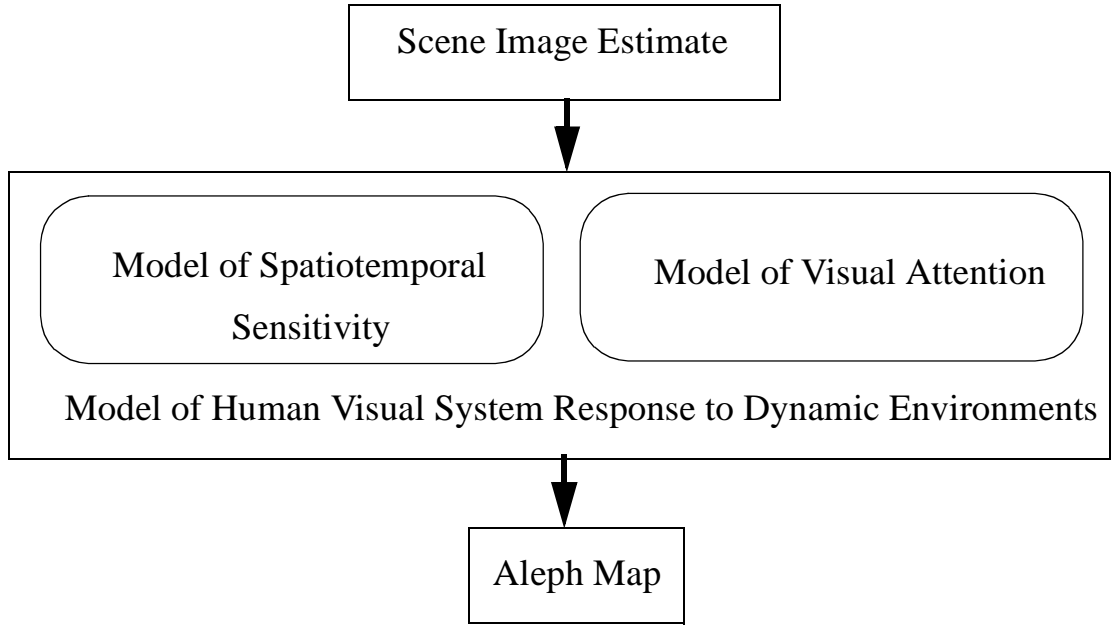

Figure 1.3: Overview of the Perceptual Model

A scene image estimate is first generated using fast graphics hardware. The estimate is processed using a computational model of visual attention and spatiotemporal sensitivity to derive the Aleph Map. The Aleph Map is then used as a perceptual oracle to guide global illumination algorithms, reducing lighting calculation times by an order of magnitude. The cost of calculating the Aleph Map is negligible.

This thesis is organized in the following way: Chapter 2 will deal with the previous work related to this thesis, and Chapter 3 will cover the background material concerning spatiotemporal sensitivity and visual attention. Chapter 4 will discuss our framework and an overview of the components required to apply the framework to global illumination. Chapter 5 will cover the implementation details of the framework and a practical augmentation of the widely used lighting solver RADIANCE. Results will be presented in Chapter 6 and we will conclude the thesis in Chapter 7 with some discussion and ideas for future work in this area.

# CHAPTER 2
# Previous Work

This chapter reviews previous work that uses perceptual techniques to speed up global illumination computation. The chapter begins with some early work that incorporated minimal perceptual elements in progressive rendering schemes and proceeds on to a review of techniques that use sophisticated models of the human visual system in progressive rendering. An excellent overview of perceptually-driven radiosity methods is given in [Prik99] and will not be repeated here.

## 2.1 Perceptually-Based Adaptive Sampling

Early work in perceptually assisted rendering was mainly concerned with the acceleration of ray tracing. These algorithms focused on anti-aliasing an image while at the same time shooting as few samples as possible. The algorithms usually sampled an image at a low density, applied some simple perceptual model to the low density image, adaptively supersampled according to the perceptual model, and then reconstructed the final image solution using some interpolating filter kernel.

Mitchell [Mitc87] was one of the first to use a simple model of the human visual system to assist a rendering algorithm. Mitchell's model of the human visual system is incomplete but was one of the first to take advantage of the decrease in the visual system's sensitivity at high spatial frequencies and contrasts. The first part of his algorithm was the generation of a Poisson-disk sampled image. The Poisson-disk sampling shifted aliasing artifacts to higher frequencies, where the visual system is less likely to notice errors. Next, his adaptive sampling technique took into account the non-linear response of the eye to changes in intensity by using a weighted contrast measure of the red, green and blue channels of the Poisson-disk sampled image. The algorithm then supersamples regions of the image for which the contrast measure exceeded a set threshold to obtain the adaptively sampled image. The adaptively sampled image is then convolved with a reconstruction filter to obtain the final rendered image.

Another paper that made use of an adaptive sampling scheme was written by Painter and Sloan [Pain89]. Painter's adaptive algorithm generated a k-D tree that partitioned the image plane into rectangular regions recursively. At each node of the tree was a variance estimate that contained the approximation error estimate for the node. The algorithm refined each node of the tree progressively until pixel level accuracy was attained. Painter and Sloan briefly mention that their algorithm could take into account the non-linear response of the human visual system to intensity variations but do not explicitly describe the implementation of a perceptual measure in their paper. Meyer and Liu [Meye92] however, extended the Painter and Sloan algorithm to take advantage of the limited sensitivity of the human visual system to color variations. In their scheme, more rays were shot at places in the scene with intensity changes rather than color changes, resulting in a moderate speedup in rendering.

Bolin and Meyer [Boli95] developed a frequency-based ray tracer that rendered directly into the frequency domain and used a more complete model of the human visual system to guide rendering than the earlier algorithms. Their technique took into account two characteristics of the human visual system: contrast sensitivity and spatial frequency response. The visual system is less likely to notice errors in regions of an image of high contrast, or high spatial frequency. Since the ray tracer operates in the frequency domain, the algorithm could choose to spawn more rays when there is low contrast or low spatial frequency. The ray tracer also spawned more rays for spatial luminance changes rather than color changes.

## 2.2 Perceptual Metrics

Another class of perceptually-based renderers use sophisticated perceptual metrics to inform the renderer to stop calculating when the lighting solution has errors that are below a threshold determined by a perceptual model of the visual system. The basic operation in these techniques is to perform a comparison between two intermediate images of a rendering and to stop the rendering whenever the measure of error between the two images is smaller than some threshold. The error between two images can be something as simple as the absolute difference between the images, or the sum of squared differences between corresponding pixels of the images. This kind of error metric is known as a physically based error metric. Another form of error metric is a perceptual error metric, in which the physical information is fed to a model of the human visual system before the error computation is performed. Figure 2.1 is a conceptual diagram of the use of a perceptual metric in progressive rendering.

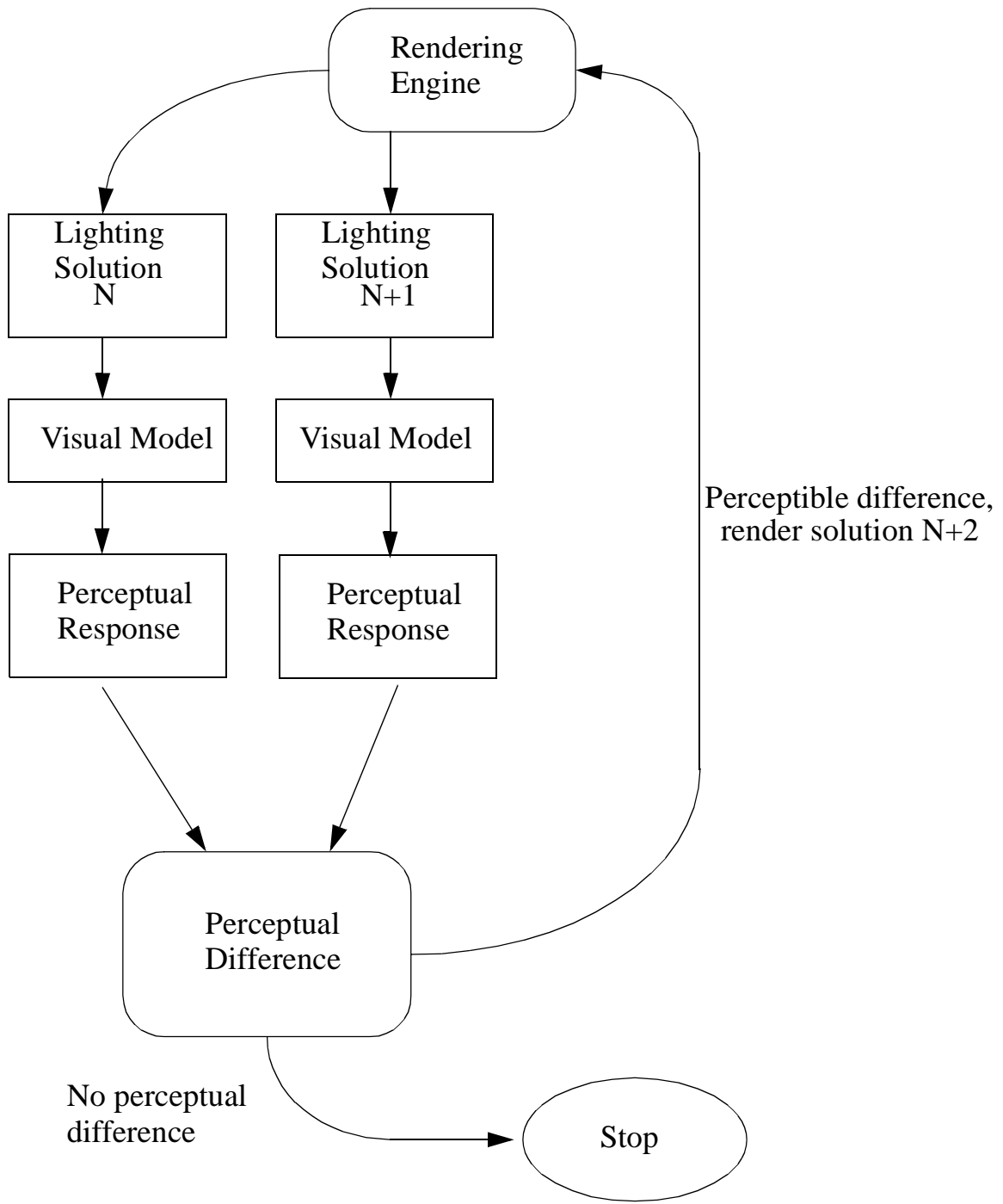

Figure 2.1: Conceptual Diagram of Perceptually-assisted Rendering

Perceptually-assisted rendering makes use of a perceptual metric to compare intermediate results of a lighting computation, directing the rendering to halt when there is no perceptual difference between the two intermediate results.

In the figure, the rendering engine produces two images at different stages of the lighting calculation. The two images are compared using the perceptual metric and if there is no perceptual difference between them, the algorithm halts and the image is said to be converged, at least in the perceptual domain. Otherwise, the rendering engine continues the lighting computation and repeats the process until the perceptual differences are small enough to meet a user specified convergence limit.

The Daly Visible Differences Predictor (VDP) [Daly93] and the Sarnoff Visual Discrimination Model (VDM) [Lubi95] are two commonly used perceptual metrics. Both metrics measure the perceptual differences between two images and calculate the difference using sophisticated models of the human visual system.

The Daly VDP, given two images, returns a map that contains the probability of detection of differences between the images. The VDP takes into account the light levels, the spatial frequency content and the orientation signal content in the images. The process begins by applying an amplitude nonlinearity function to the luminance channel of the image. This models the retinal response to the image luminance. In the next stage, the *Contrast Sensitivity Function* is used to determine the visual sensitivity to spatial patterns in the retinal response image. The Contrast Sensitivity Function is an experimentally derived equation that quantifies the human visual sensitivity to spatial frequencies and will be covered in detail in the next section of this chapter. The CSF was used in the VDP as a normalization factor for the subsequent detection mechanisms. There are four detection mechanisms in the Daly VDP: the spatial frequency hierarchy, the masking function, the psychometric function and probability summation [Daly93]. In the spatial frequency hierarchy stage, the image is decomposed into several frequency sub-bands and orientation bands, and then weighted by the CSF and the masking function. The masking function models the decrease in sensitivity to a signal in the presence of another signal in the same frequency subband. The product of the CSF and the masking function is known as the threshold elevation factor. This elevation factor is used to weight the spatial frequency and orientation signals. The two images being compared are identically subjected to the signal processing described above, and the differences between the signals are applied to a psychometric function which converts a contrast difference into a probability of detection. The probabilities for each spatial frequency and orientation channel are combined to derive a final per pixel probability of detection value. Figure 2.2 provides a graphical overview of the Daly VDP.

The Sarnoff Visual Discrimination Model [Lubi95] is another well designed perceptual metric that is used for determining the perceptual differences between two images. The Sarnoff VDM returns the map of Just-Noticeable-Differences (JNDs) between the images and is an image space algorithm, similar to the Daly VDP. The Sarnoff VDM begins by convolving the image with a point spread function that models the effect of the optics of the eye on the image. The processed image is then re-sampled according to foveal eccentricity to model the decrease in spatial resolution away from the foveal regions of the retina. Next, the band-pass contrast responses are obtained by decomposing the image into a contrast pyramid using the technique of Burt and Adelson [Burt83]. Each band-pass contrast response level of the pyramid is then processed using the steerable filters of Freeman and Adelson [Free91] in order to extract orientation information from each level. The output of the orientation filtering is then summed and weighted by the CSF and passed through a non-linear transform to model the changes in sensitivity with spatial frequency and contrast respectively. The two images to

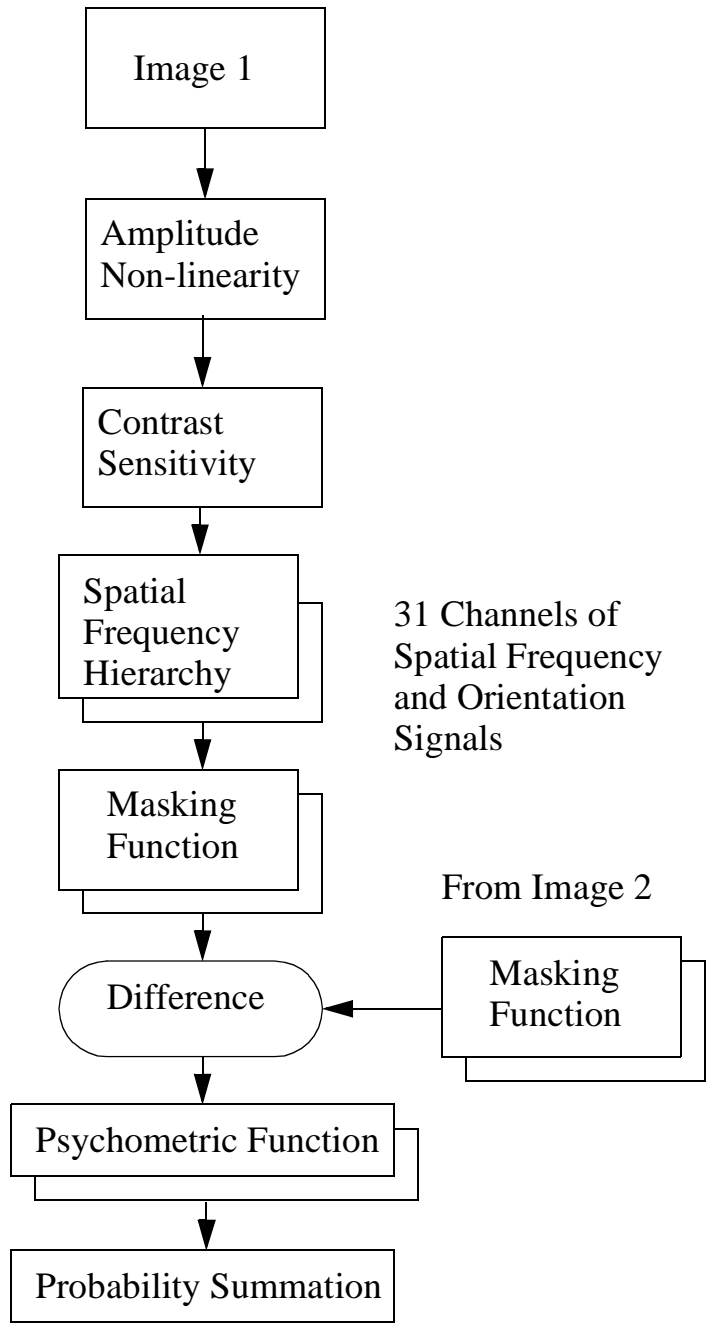

Figure 2.2: Daly VDP Overview

The process by which two images are compared for perceptual differences in the Daly VDP. The images are decomposed into a series of spatial frequency and orientation signals, weighted by the CSF and the masking function. The difference of signals is taken between the two images and converted to probabilities using the psychometric function and summed to derive a single per pixel probability of detection.

be compared are subjected to identical processing as described above and a distance measure is computed by taking the difference in responses for each channel processed and summing them together to obtain the JND map of the two images. Figure 2.3 provides a graphical overview of the Sarnoff VDM.

Li, et. al., [Li98] provide an excellent comparison of the Daly and Sarnoff perceptual metrics in their paper.

A comprehensive model of visual masking was developed by Ferwerda et. al. [Ferw97] that can be used to predict how the presence of one visual pattern affects the detectability of another visual pattern when they are superimposed over each other. Although it is not a perceptual metric per se, it can be used to predict the effects of texture mapping on masking tesselation or the effect of geometric complexity on masking rendering artifacts.

## 2.3 Applications of Perceptual Metrics

Bolin and Meyer [Boli98] and Myszkowski [Mysz98] relied on the use of sophisticated perceptual metrics to estimate perceptual differences between two images to determine the perceived quality at an intermediate stage of a lighting computation. Based on perceptual quality, they determined the perceptual convergence of the solution and used it as a stopping condition in their global illumination algorithm. These metrics perform signal processing on the two images to be compared, mimicking the response of the visual cortex to spatial frequency patterns and calculating a perceptual distance between the two images. Myskowski uses the Daly Visible Differences Predictor [Daly93] to determine the stopping condition of rendering by comparing two images at different stages of the lighting solution. Bolin and Meyer used a computationally efficient and simplified variant of the Sarnoff Visual

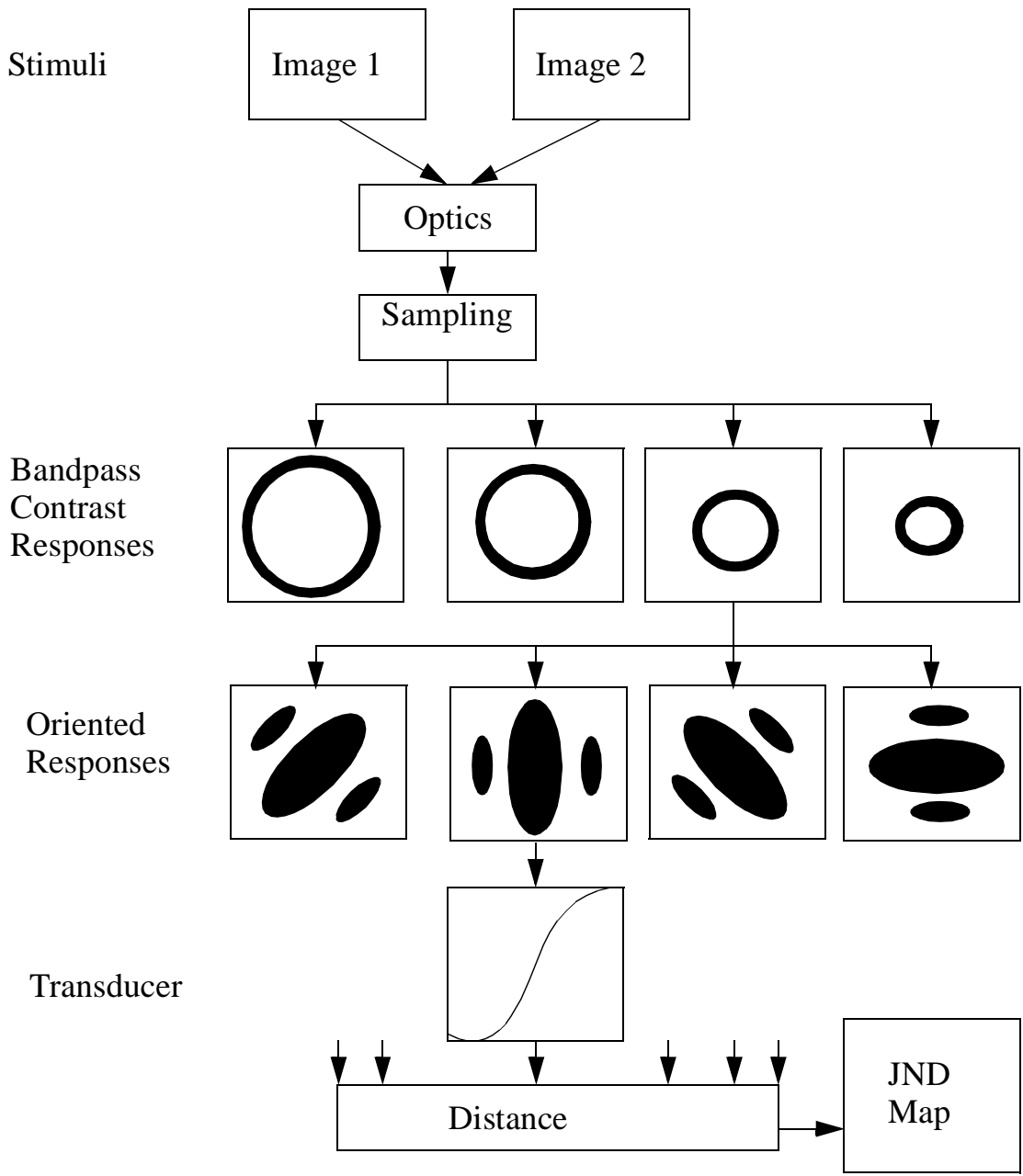

Figure 2.3: Sarnoff Visual Discrimination Model (adapted from [Lubi95]).

The Sarnoff VDM processes two images using a series of spatial decomposition filters and perceptual models to compute a Just-Noticeable-Difference (JND) map. This map is a measure in perceptual space of the perceptual "distance" between two images, and can be converted into a probability of detection value in a similar manner as the Daly VDP.

Discrimination Model [Lubi95] on an upper bound and a lower bound pair of images to determine the stopping condition in their bounded-error, perceptually-guided algorithm. Both algorithms required repeated applications of the perceptual error metric at intermediate stages of a lighting solution. The repeated application of the metric added substantial overhead to the rendering algorithm.

Ramasubramanian, et al., [Rama99] reduced the cost of such metrics by decoupling the expensive spatial frequency component evaluation from the perceptual metric computation. They reasoned that the spatial frequency content of the scene does not change significantly during the global illumination computation step and, hence, precomputed this information from a cheaper estimate of the scene image. They reused the spatial frequency information during the evaluation of the perceptual metric, without having to recalculate it at every iteration of the global illumination computation. They carried out this pre-computation from the direct illumination solution of the scene. Their technique does not take into account any sensitivity loss due to motion and hence is not well suited for use in dynamic environments.

Myskowski, et al., [Mysz99] addressed the perceptual issues relevant to rendering dynamic environments. They incorporated spatiotemporal sensitivity of the HVS into the Daly VDP [Daly93] to create a perceptually-based Animation Quality Metric (AQM) and used it in conjunction with image-based rendering techniques [McMi97] to accelerate the rendering of a key-frame based animation sequence. Myskowski's framework assumed that the eye tracks all objects in a scene. The tracking ability of the eye is very important in the consideration of spatiotemporal sensitivity [Daly98]. Perceptually-based rendering algorithms which ignore this ability of the eye can introduce perceptible error in visually salient areas of the scene. On the other hand, the most conservative approach of indiscriminate tracking of all the objects of a scene, as taken by Myskowski's algorithm, effectively reduces a dynamic scene to a static scene, thus negating the benefits of spatiotemporally-based perceptual acceleration. The use of AQM during global illumination computation will also add substantial overhead to the rendering process.

## 2.4 Our Approach

Our technique combines the best of existing algorithms by developing a spatiotemporal error tolerance map, the Aleph Map, that takes into account not only spatial information but temporal as well. It is quickly precomputed from frame estimates of the animation to be rendered, and are estimates that capture spatial frequency and motion correctly. We make use of fast graphics hardware to obtain the Aleph Map quickly and efficiently. The map is correct because it incorporates a model of visual attention in order to compensate for the tracking ability of the eye.

The Aleph Map can be adapted for use as a physically-based error metric, or as in our application, as an oracle that guides perceptual rendering without the use of an expensive comparison operator. By using a perceptual oracle instead of a metric, we incur negligible overhead while rendering. Our approach will address the issue of overhead when using perceptual techniques, as well as correctly accounting for the tracking ability of the eye.

The next chapter will develop the background material required to understand the construction of the Aleph Map. It will discuss spatiotemporal sensitivity in the context of perceptually-assisted rendering. The chapter will also expound on the importance of visual attention with regards to dynamic scenes and the tracking behavior of the eye in such scenes.

# CHAPTER 3
# Background

*The cosmic [microwave] background radiation suffuses the entire universe -*
*Carroll & Ostlie, An Introduction to Modern Astrophysics*

This chapter covers the background material relevant to this thesis. The first part will review the spatiotemporal contrast sensitivity of the human visual system and the second part will address the attention mechanisms of the visual system.

## 3.1 Spatiotemporal Contrast Sensitivity

### 3.1.1 Contrast Sensitivity

The sensitivity of the Human Visual System (HVS) changes with the spatial frequency content of the viewing scene. This sensitivity is psychophysically derived by measuring the threshold contrast for viewing sine wave gratings at various frequencies [Camp68]. The Contrast Sensitivity Function (CSF) is the inverse of this measured threshold contrast, and is a measure of the sensitivity of the HVS towards static spatial frequency patterns. This CSF function peaks between 4-5 cycles per degree (cpd) and falls rapidly at higher frequencies. The reduced sensitivity of the HVS to high frequency patterns allows the visual system to tolerate greater error in high frequency areas of rendered scenes and has been exploited extensively by [Boli95], [Boli98] , [Mysz98], [Mysz99] and [Rama99] in the rendering of static scenes containing areas of high frequency texture patterns and geometric complexity.

### 3.1.2 Temporal Effects

The HVS varies in sensitivity not only with spatial frequency but also with motion. Kelly [Kell79] has studied this effect by measuring threshold contrast for viewing travelling sine waves. Kelly's experiment used a special technique to stabilize the retinal image during measurements and therefore his models use the retinal velocity, the velocity of the target stimulus with respect to the retina. Figure 3.1 summarizes these measurements.

From the figure, we can see that the contrast sensitivity changes significantly with the retinal velocity. Above the retinal velocity of 0.15 deg/sec, the peak sensitivity drops and the entire curve shifts to the left. This shift implies that waveforms of higher frequency become increasingly difficult to discern as the velocity increases. At retinal velocities below 0.15 deg/sec the whole sensitivity curve drops significantly. Speeds below 0.15 deg/sec are artificial as the eye naturally drifts about randomly even when it is staring fixatedly at a single point. The measurements also show that the sensitivity function obtained at the retinal velocity of 0.15 deg/sec matched with the static CSF function described earlier. This agrees with the fact that the drift velocity of a fixated eye is about 0.15 deg/sec, and must be taken into account when using Kelly's measurement results in real world applications.

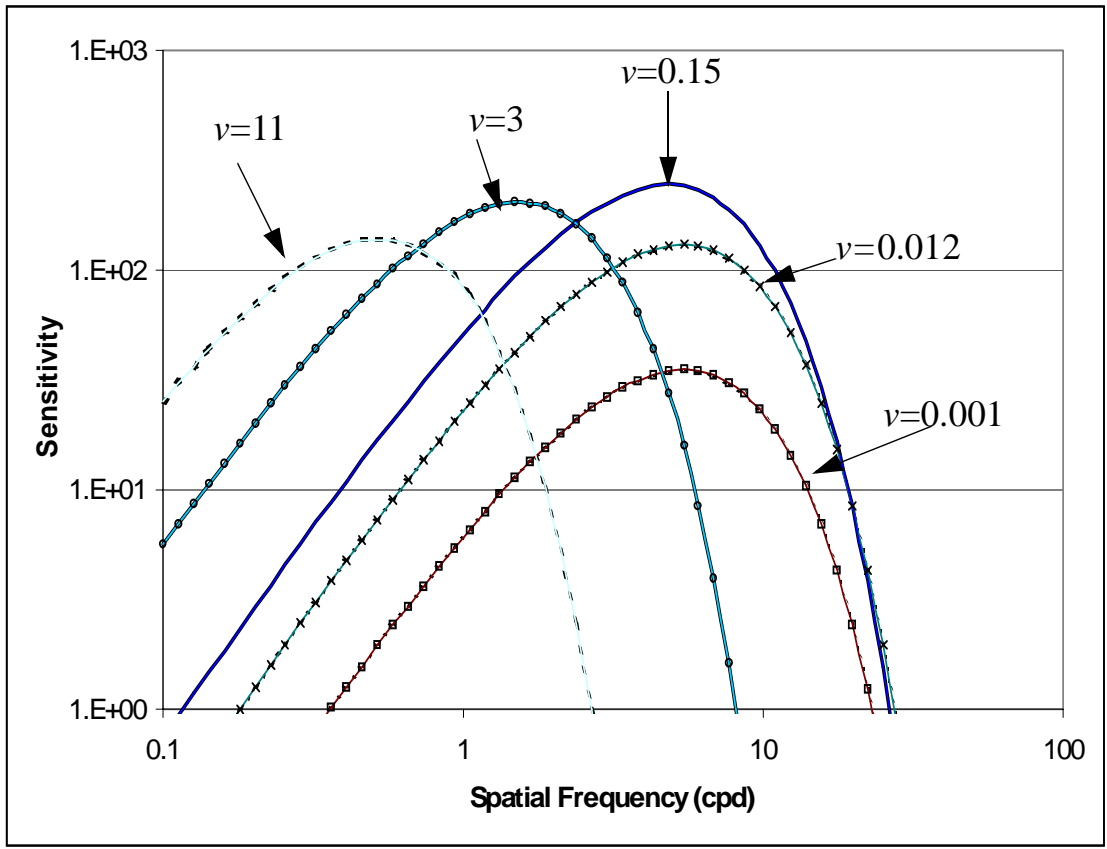

Figure 3.1: Velocity Dependent CSF

Plotted from an equation empirically derived from Kelly's sensitivity measurements [Daly98]. The velocities $v$ are measured in degrees/second.

### 3.1.3 Eye Movements

The loss of sensitivity to high frequency spatial patterns in motion gives an opportunity to extend existing perceptually-based rendering techniques from static environments to dynamic environments. The eye, however, is able to track objects in motion to keep objects of interest in the foveal region where spatial sensitivity is at its highest. This tracking capability of the eye, also known as smooth pursuit, reduces the retinal velocity of the tracked objects and thus compensates for the loss of sensitivity due to motion. Figure 3.2 is a chart graphically portraying the tracking capabilities of the eye as the speed of a target increases.

Measurements by Daly [Daly98] have shown that the eye can track targets cleanly at speeds up to 80 deg/sec. Beyond this speed, the eye is no longer able to track perfectly. The results of such measurements are shown in Figure 3.2. The open circles in Figure 3.2 show the velocity of the eye of an observer in a target tracking experiment. The measured tracking velocity is on the vertical axis while the actual target velocity is on the horizontal axis. The solid line in Figure 3.2 represents a model of the eye's smooth pursuit motion.

Evidently, it is crucial that we compensate for smooth pursuit movements of the eye when calculating spatiotemporal sensitivity. The following equation describes a motion compensation heuristic proposed by Daly [Daly98]:

$$v_R = v_I - min(0.82v_I + v_{Min}, v_{Max}) \tag{3.1}$$

where $v_R$ is the compensated retinal velocity, $v_I$ is the physical velocity, $v_{Min}$ is 0.15 deg/sec (the drift velocity of the eye), $v_{Max}$ is 80 deg/sec (which is the maximum velocity that the eye can track efficiently). The value 0.82 accounts for Daly's data fitting that indicates the eye tracks all objects in the visual field with an efficiency of 82%. The solid line in Figure 3.2 was con-

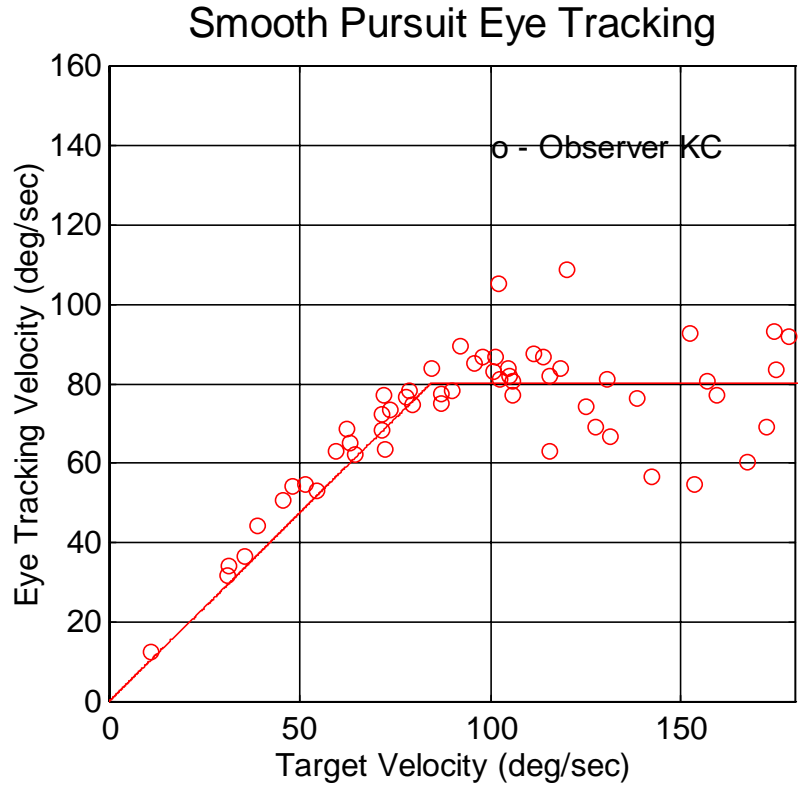

Figure 3.2: Smooth Pursuit Behavior of the Eye

The eye can track targets reliably up to a speed of 80.0 deg/sec beyond which tracking is erratic. Reproduced from Daly [Daly98].

structed using this fit. Use of this heuristic would imply only a marginal improvement of efficiency in extending perceptual rendering algorithms for dynamic environments, but our method offers an order of magnitude improvement.

## 3.2 Visual Attention and Saliency

Though the eye's smooth pursuit behavior can compensate for the motion of the moving objects in its focus of attention, not every moving object in the world is the object of one's attention. The pioneering work of Yarbus [Yarb67] shows that even under static viewing conditions not every object in the viewing field captures visual attention. If we can predict the focus of attention, then other less important areas may have much larger error tolerances allowing us to save calculation time on those areas. To accomplish this, we need a model of visual attention which will correctly identify the possible areas of visual interest.

Visual attention is the process of selecting a portion of the available visual information for localization, identification and understanding of objects in an environment. It allows the visual system to process visual input preferentially by shifting attention about an image, giving more attention to salient locations and less attention to unimportant regions. The scan path of the eye is thus strongly affected by visual attention. In recent years, considerable efforts have been devoted to understanding the mechanism driving visual attention. Contributors to the field include Yarbus [Yarb67], Yantis [Yant96], Tsotsos, et al. [Tsot95], Koch and Ullman [Koch85], Niebur & Koch [Nieb98].

Two general processes significantly influence visual attention, called bottom-up and top-down processes. The bottom-up process is purely stimulus driven. A few examples of such stimuli are: a candle burning in a dark room; a red ball among a large number of blue balls; or sudden motions. In all these cases the conspicuous visual stimulus captures attention automatically without volitional control. The top-down process, on the other hand, is a directed volitional process of focusing attention on one or more objects which are relevant to the observer's goal. Such goals may include looking for street signs or searching for a target in a computer game. Though the attention drawn due to conspicuity may be deliberately ignored because of irrelevance to the goal at hand, in most cases, the bottom-up process is thought to provide the con-text over which the top-down process operates. Thus, the bottom-up process is fundamental to the visual attention.

We disregard the top-down component in favor of a more general and automated bottom-up approach. In doing so, we would be ignoring non-stimulus cues such as a "look over there" command given by the narrator of a scene or shifts of attention due to familiarity. Moreover, a task driven top-down regime can always be added later, if needed, with the use of supervised learning [Itti99a].

Itti, Koch and Niebur [Itti00][Itti99a][Itti99b][Itti98] have provided a computational model to this bottom-up approach to visual attention. The model is built on a biologically plausible architecture proposed by Koch and Ullman [Koch85] and by Niebur and Koch [Nieb98]. Figure 6 graphically illustrates the model of visual attention. The figure, which illustrates an abridged version of the process, is shown for the achromatic intensity channel. In the figure, feature maps, which represent zones of interest in a specific channel at a specific scale, are combined to get a summary of interesting areas in a specified channel at all scale levels called the conspicuity map. The conspicuity maps of the channels of intensity, color, orientation and motion are combined to obtain the saliency map. Bright regions on the maps denote areas of interest to the visual system.

The computational architecture of this model is largely a set of center-surround linear operations that mimic the biological functions of the retina, lateral geniculate nucleus and primary visual cortex [Leve91]. The center surround effect makes the visual system highly sensitive to features such as edges, abrupt changes in color and sudden movements. This model generates feature maps, using center surround mechanisms, for visually important channels such as intensity, color and orientation. A feature map can be con-sidered to represent the conspicuity at different spatial scales. Each of these features for each of these channels is computed at multiple scales and then processed with an operator, N(.), that mimics the lateral inhibition effect. That is, features that are similar and near each other cancel each other out. Feature maps that have outstanding features are emphasized while feature maps which have competing features or no outstanding features are suppressed. For example, a single white square in a dark background would be emphasized, but a checkerboard pattern would be suppressed. The sum of the feature maps for each channel after they have been processed for lateral inhibition results in a conspicuity map. The conspicuity map for each channel are processed themselves for lateral inhibition and then summed together to obtain a single saliency map that quantifies visual attention. The model of Itti, et al., has been tested with real world scenes and has been found to be effective [Itti00].

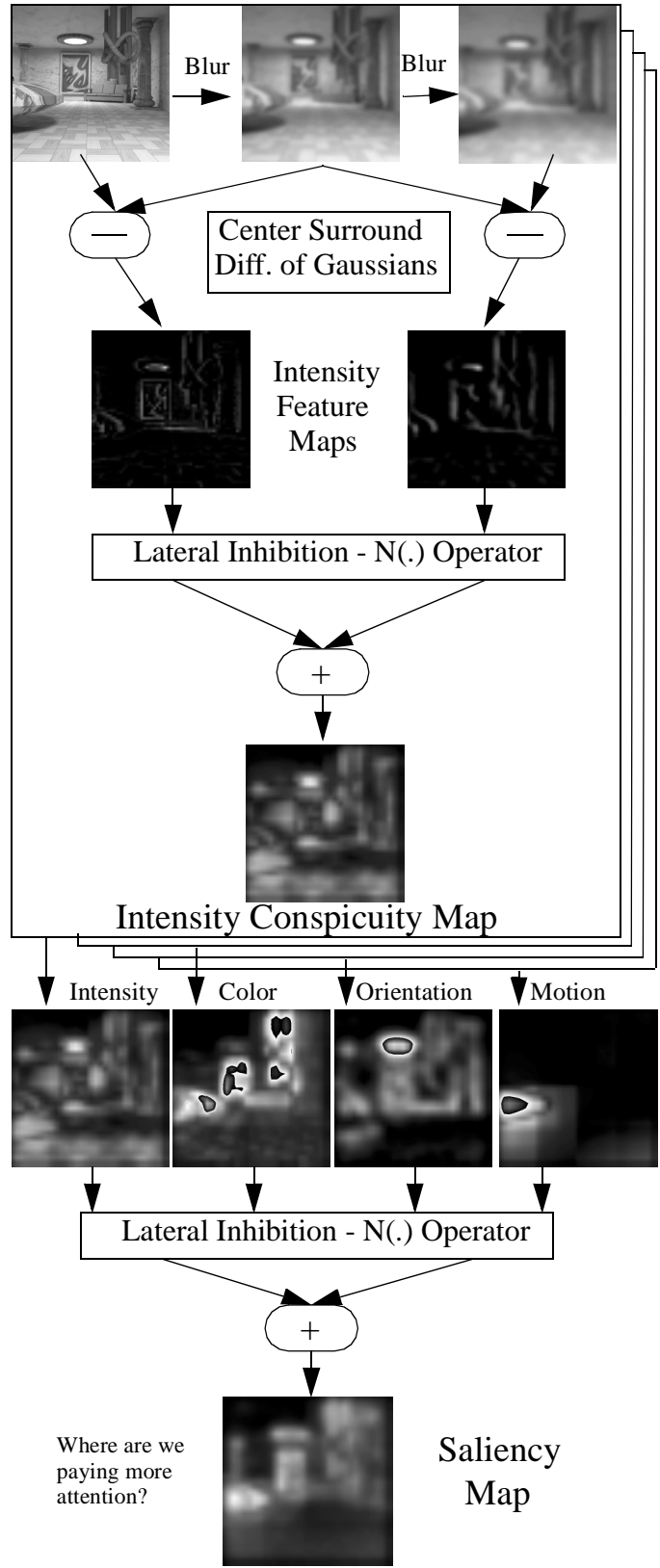

Figure 3.3: Outline of the Computational Model of Visual Attention

The model of Itti, Koch and Niebur does not include motion as a conspicuity channel. We include motion as an additional conspicuity channel in our implementation. The next chapter describes the an overview of the process of obtaining the Aleph map by building on the knowledge presented here.

# CHAPTER 4
# Framework

*B'roshyth bara Elohiym et ha-shomayim v'et ha-aretz. (In the beginning God created the heavens and the Earth). - first words in the Torah.*

This chapter establishes the framework for calculating the Aleph Map. Our process begins with a rapid image estimate of the scene. This image estimate serves both to identify areas where spatiotemporal sensitivity is low and also to locate areas where an observer will be most likely to look. Such an image may be quickly generated using an Open GL rendering, or a ray traced rendering of the scene with only direct lighting. We have typically used Open GL to render estimates (one for each frame of the estimation) for our work and use the estimate only for the computation of the Aleph Map and not for the actual global illumination calculation.

Our computation proceeds in four major steps: 1) motion estimation, 2) spatial frequency estimation, 3) saliency estimation and 4) computing the Aleph Map. We will discuss each of these steps in detail in the following section. Figure 4.1 depicts an overview of the process.

Motion estimation is used to calculate spatiotemporal sensitivity as well as saliency. The motion estimate can be deduced from the way the geometry is transformed or from how much pixels move from one frame or another. The spatial frequency content of the scene is computed and used in both the spatiotemporal sensitivity and the saliency calculation. Finally, the motion,

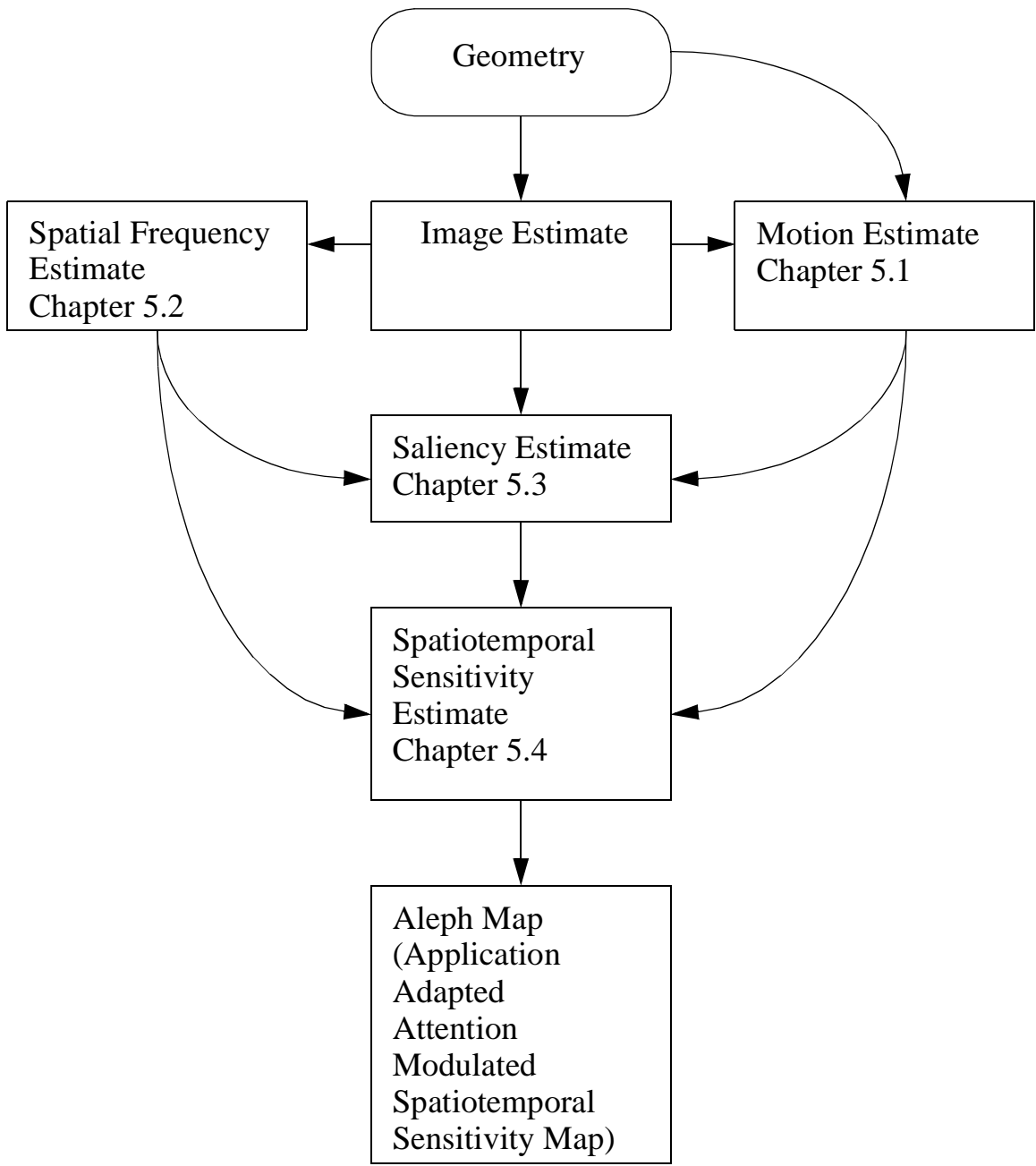

Figure 4.1: Framework for Aleph Map Computation

An overview of the process to compute the Aleph Map is shown. The process begins with an Image Estimate which is used to compute spatial frequency, motion and saliency. The information is then combined to obtain the Aleph Map, a representation of spatiotemporal sensitivity that also takes into account the attention mechanisms of the visual system.

spatial frequency and saliency are all combined to obtain the spatiotemporal sensitivity which is adapted to a particular application. This application-adapted spatiotemporal sensitivity is called the Aleph Map. We will now proceed with the implementation details in calculating the Aleph Map.

# CHAPTER 5

# Implementation

*V'omer Elohiym hayah ohr v'hayah ohr. (And God said, "Let there be light," and there was light) - Biblical Universe Implementation*

This chapter goes into detail on the computation steps required to derive the Aleph Map. The prerequisite for the computation is a series of consecutive image frames that are estimates of the animation to be rendered. These images are used to estimate motion, spatial frequency and saliency, the ingredients needed for calculating spatiotemporal sensitivity. An example of two kinds of image estimates are shown in Figure 5.1. As can be seen in the figure, the Open GL image estimate captures the relevant spatial frequency and color information in a scene prior to rendering, while the ray traced image estimate captures shadow effects as well. However, the ray traced image can be expensive to calculate in comparison to the Open GL image estimate.

Before they are used, the image estimates are converted from RGB into $AC_1C_2$ opponent color space, using the transformation matrices given in [Boli95]. The color space conversion facilitates the computation of visual saliency. We use the following notation in our description. A capital letter such as 'A' or '$C_1$' or '$C_2$' denotes a channel and a number in parenthesis denotes the level of scale. Thus, 'A(0)' would correspond to the finest scale

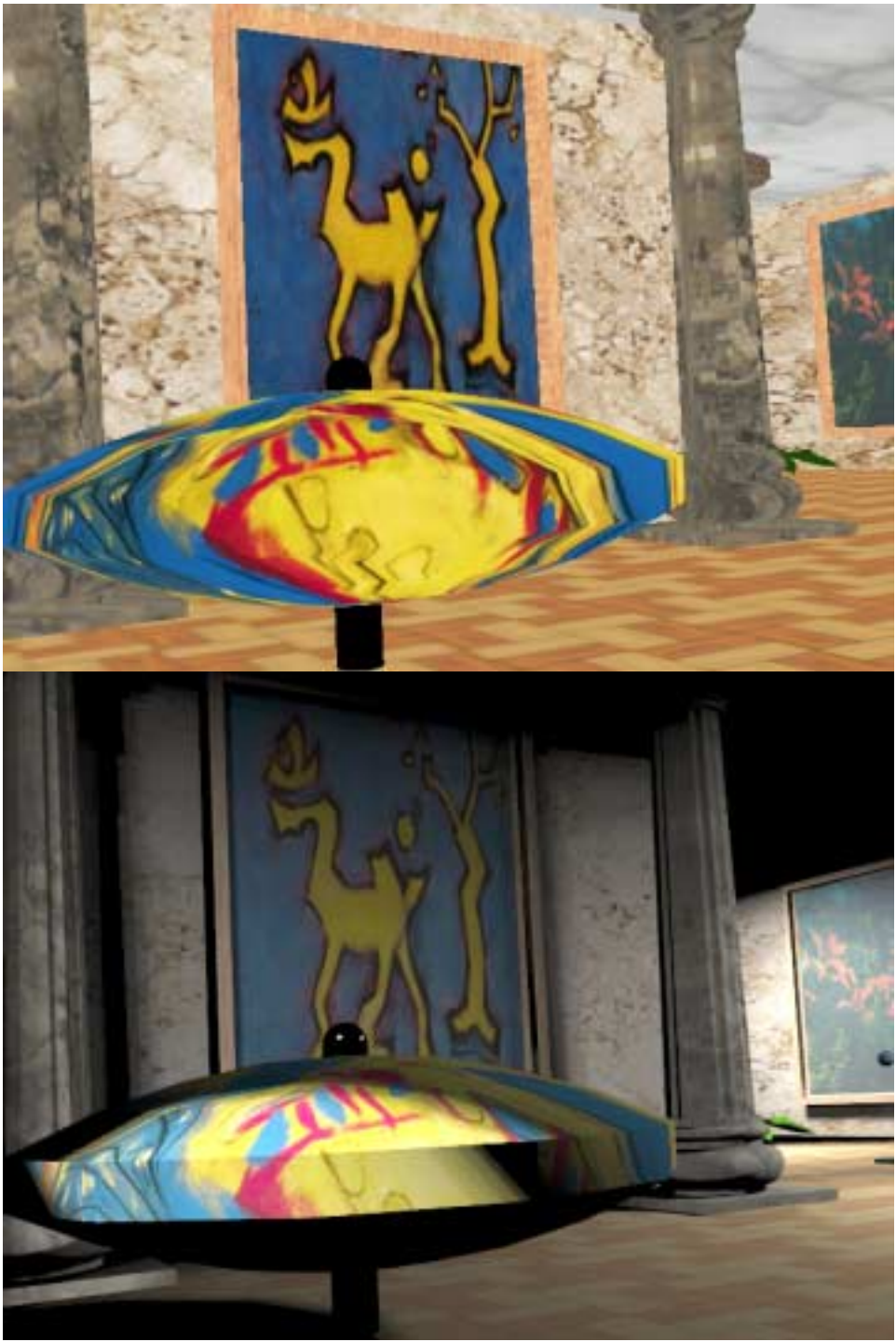

(a)

Open GL

(b)

Ray
Traced

Figure 5.1: Image Estimates

Image (a) is an Open GL Image estimate (generated in less than a second)
and (b), a ray traced image estimate (generated in 5 minutes).

of a multiscale decomposition of the achromatic channel of the $AC_1C_2$ color space. For conciseness, a per-pixel operation, e.g. A(x,y), is implied.

## 5.1 Motion Estimation

The velocity of pixels in the image plane is needed to estimate both spatiotemporal sensitivity and visual attention. We implemented two different techniques to estimate image plane velocity. One makes use of the image estimate alone and the other makes use of additional information such as geometry and knowledge of the transformations used for movement. The latter model is appropriate for model-based image synthesis applications while the former can be used even when only the image is available, as in image-based rendering. In both of these techniques, the goal is first to estimate displacements of pixels $\Delta P(x,y)$ from one frame to another, and then to compute the image velocity from these pixel displacements, using frame rate and pixel density information.

### 5.1.1 Image-based Motion Estimation

Image-based motion estimation is useful when the geometry and their transforms are not directly available, such as when the source data consists of only image frames. In such cases, the pixel displacement can be obtained by tracking the motion of pixels from one frame to another. Figure 5.2 shows how image-based motion estimation works in general.

In the figure, points in Frame N are tracked in Frame N+1 by searching for their new locations around the neighborhood of the points. This is usually achieved by minimizing the sum of squared differences between the region

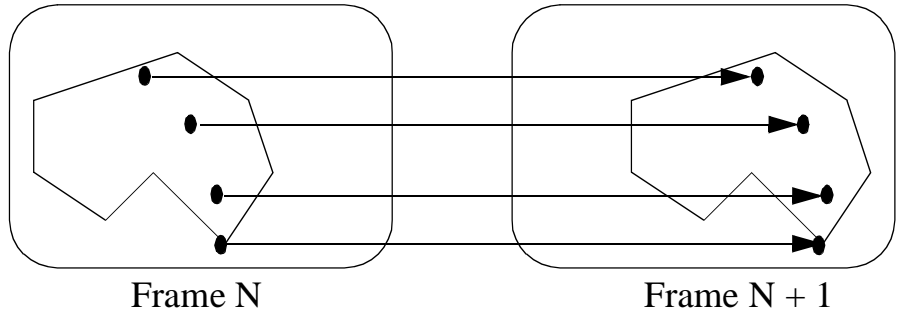

Figure 5.2: Image-based Motion Estimation

Each pixel in Frame N is tracked using pattern matching algorithms to their new locations in Frame N+1. The resulting displacement information, combined with knowledge of frame rate and pixel density, results in the actual speed of the pixels across the image plane.

around the pixel in Frame N and the target search region in Frame N+1. In effect, we are calculating the pixel displacements which, when applied to Frame N, will give an image that has the smallest differences from Frame N+1. We now describe a faster, hierarchical variant of the above scheme that uses a logical operator instead of pixel differences to search for pixel displacements.

In this hierarchical, image-based motion estimation technique, the achromatic 'A' channels of two consecutive image frames are decomposed into multiscale Gaussian pyramids using the filtering method proposed by Burt and Adelson [Burt83]. The Gaussian filtered images are then processed by the census transform [Zabi94], a local transform that is used to improve the robustness of motion estimation. The census transform generates a bitstring for each pixel that is a summary of the local spatial structure around the pixel. The bits in the bitstring correspond to the neighboring pixels of the pixel under consideration. The bit is set to 0 if the neighboring pixel is of lower intensity than the pixel under consideration. Otherwise, it is set to 1. Figure 5.3 illustrates the Census Transform.

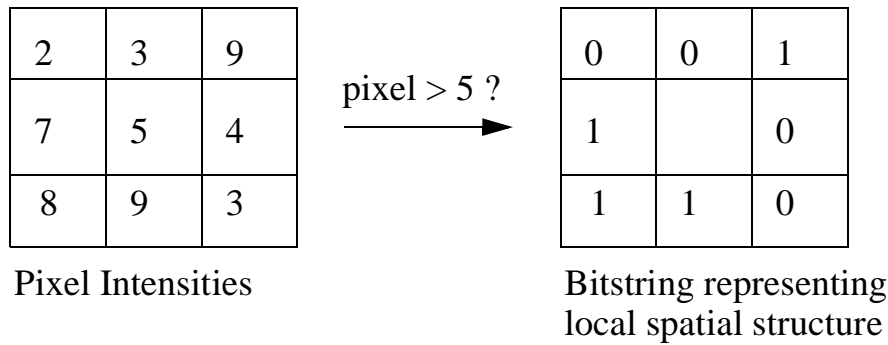

Pixel Intensities

Bitstring representing local spatial structure

Figure 5.3: Census Transform

The intensity values of the central pixel '5' and its neighbors are compared. Every neighbor with an intensity less than 5 will be given a 0, otherwise it is given a 1. The central pixel ('5') after undergoing the census transform will contain the bitstring '00110110' which represents the spatial structure around the pixel.

In the example given above, the pixel with value '5' and its surrounding neighbors are shown. The census transform replaces the pixel value '5' with the bitstring '00110110' representing the relative intensity changes in its neighborhood. Thus, the census transform represents the spatial structure of the surrounding pixels at a particular point. Performing the census transform allows us to find correspondences in the two images by capturing both intensity and local spatial structure. It also makes motion estimation effective against exposure variations between frames (if a real world photograph was used). Comparisons can then be made between regions of census transformed images by calculating the minimum Hamming distance between two bit strings being compared. The Hamming distance of two bit strings is defined as the number of bits that are different between the two strings and can be implemented efficiently by eXclusive-ORing the two strings together and counting the number of '1' bits. For example, the Hamming distance of "1110" and "1011" is 2 because the second and fourth bits are different. Counting the "1"s after an exclusive-or operation gives us the Hamming distance of two bit strings directly.

The census transform is now applied to levels A(0,1,2) of the achromatic Gaussian pyramid. The three levels were picked as a trade-off between computational efficiency and accuracy. An exhaustive search would be most accurate but slow, and a hierarchical search would be fast but inaccurate. To take advantage of speed and accuracy, we use both kinds of searches at different levels of the pyramid. We perform an exhaustive search on the census transformed A(2), which is cheap due to its reduced size (128x128 for a 512x512 image), to figure out how far pixels have moved between frames. In our implementation, a pixel in level A(2) of frame N is searched for in level A(2) of frame N+1 with a search window of radius 8. The pixel is assumed to have moved at most 8 pixels to the left, right, up or down from its initial location in frame N. This search operation is carried out by finding the pixel displacement that minimizes the Hamming distance between the Frame N pixel and the Frame N+1 pixel. Figure 5.4 is a diagram showing how an exhaustive search is performed. For simplicity, only per pixel Hamming distance comparisons are shown. In the actual implementation, the Hamming distance is computed as the sum of Hamming distances in a 3x3 region around the initial and target pixels, as it was found to give a better displacement result.

Subsequently, the displacement information is propagated to level 1 and a three-step search heuristic (page 104 of [Teka95]) is used to refine displacement positions iteratively. Figure 5.5 shows the progress of a three-step search routine. The three-step heuristic is a search pattern that begins with a large search radius that reduces up to three times until a likely match is

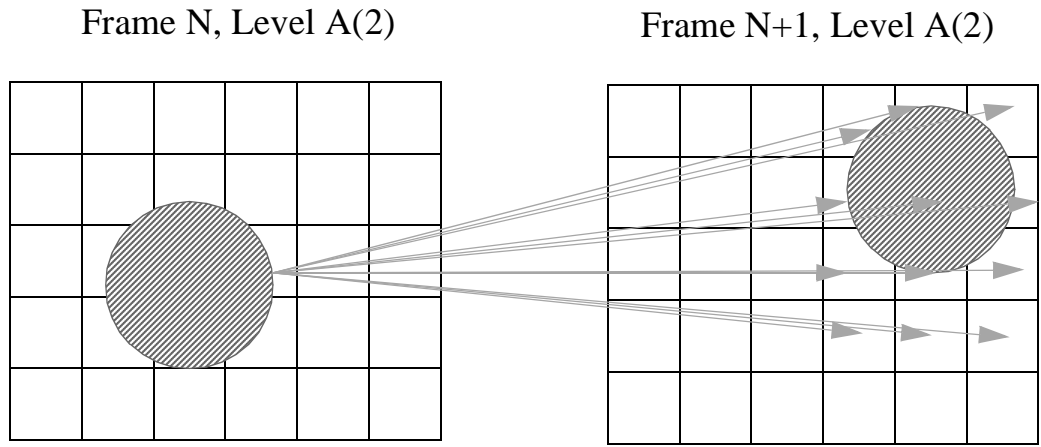

Frame N, Level A(2)  Frame N+1, Level A(2)

Figure 5.4: Exhaustive Search

In the exhaustive search algorithm, a pixel from Frame N is searched for a match in Frame N+1 by applying the Hamming distance operator to each pixel over a search region until the pixel in Frame N+1 with the smallest Hamming distance is found (dark arrow). In this diagram the search region is a 3x4 pixel window. In our implementation, the search region is 17x17 pixels wide (corresponding to a radius of 8 pixels).

found. In each step of the search, 9 pixels, including the center pixel, is checked for the closest match using the Hamming distance metric to compare the census transformed pixels. The results of level 1 is propagated to level 0 and a three-step search again conducted to get our final pixel displacement value. Each three-step search operation will look for matches in a neighborhood of radius 7 (4+2+1) around the original pixel location. Since the search is performed in a hierarchical manner, the total search radius at the resolution of the actual image (level 0) is 7+2*(7+2*8)= 53 pixels. The values come from a radius of 8 from the exhaustive search and a radius of 7 from the

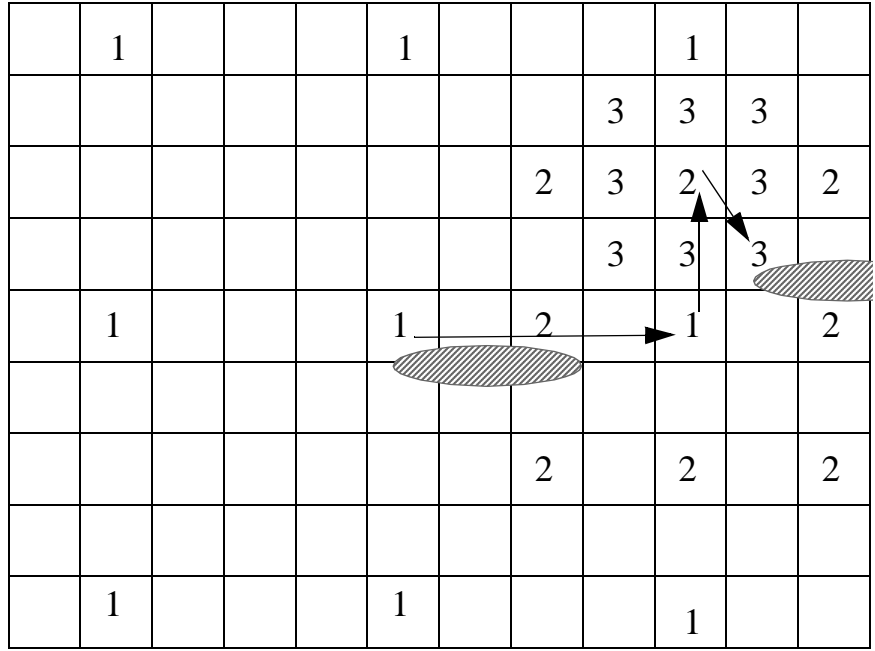

Figure 5.5: Three-step Search

The search pattern for the Three-step Search heuristic is shown. The search begins at the locations marked '1' and the most likely candidate's neighborhood is searched in a recursive manner with a shrinking neighborhood '2', '3' until the final match is found.

three-step searches. Each level is finer that the previous by a factor of two, which is why the effective radius is doubled when the displacement information is propagated to a finer level of the pyramid. The search radius is sufficient to capture the displacement of all but the fastest moving objects. Should a larger search radius be desired, one may opt to use a larger search radius when performing the exhaustive search step. The resulting displacement information will then be propagated to the finer levels of the pyramid. In effect, the pixel displacement we want, $\Delta P(x,y)$, is the displacement that minimizes the Hamming Distance between a pixel in Frame N, $P(x,y)$, and another pixel in Frame N+1, $P(x,y)+\Delta P(x,y)$.

One drawback of using an image-based technique is that the algorithms cannot calculate pixel disparities across regions of uniform color. However, it can be used in applications that do not have geometry information readily available. The following model-based motion estimation techniques is unaffected by the lack of textures and is less noisy than image-based techniques, but requires knowledge of the underlying geometry of the scene and its transformations.

## 5.1.2 Model-based Motion Estimation

When we know the geometry and transformations of each object in the scene, we can use model-based motion estimation (Agrawala, et. al. [Agra95]). In this motion estimation technique, no searching is performed. Instead, the pixel displacements are calculated by direct projection from the viewing plane (frame N), to the object, and then to the next viewing plane position (frame N+1). Thus, the running time of this motion estimation technique is proportional to the number of polygons in the geometrical database.

We begin by obtaining an object identifier and point of intersection on the object for every pixel in frame N, using either ray casting or using OpenGL hardware projection.

When ray casting is used, each pixel in frame N is projected from the camera's center of projection onto objects in world space and the Q(u,v,t) parametric coordinates of the object intersection is recorded. OpenGL may also be used to estimate motion via a two pass technique [Hanr90]. In the first pass, an Identifier (ID) Map is created by rendering each primitive in the scene with a unique color. This enables us to determine the objects struck by the projected pixels. In the second pass, a UV Map is generated that contains the barycentric u and v coordinates of each triangle. This UV map is computed by rendering each primitive with a texture map that has a red ramp in the horizontal direction and a green ramp in the vertical direction. UV coordinates are recovered from the color channels when the UV Map is drawn. Using the ID Map and the UV Map, we know the approximate projected location of a pixel onto an object in world space. With either technique, we now know for every pixel in the viewing plane, P(x,y), the projected intersection point with each object in frame N, Q(u,v,N).

Next, we advance to Frame N+1, and apply the appropriate motion transformations to Q(u,v,N), and project each point Q(u,v,N+1) onto the viewing plane corresponding to the (N+1)th frame to obtain P(x',y'). The distance of pixel movement is the displacement needed for calculating the image velocity. Figure 5.6 illustrates the model-based motion estimation procedure.

In the figure, each pixel P(x,y) in the viewing plane in Frame N is tracked to its intersection point in object space, Q(u,v,N). For static objects, Q(u,v,N)=Q(u,v,N+1). For moving objects, like the moving ball shown, the appropriate motion transformation is applied to the object and to obtain Q(u,v,N+1). The point Q(u,v,N+1) is reprojected onto the new viewing plane for Frame N+1 to location P(x',y'). The reprojection takes into account motion due to camera movements. The pixel displacement is the difference between the old and new locations $\Delta P(x,y)=P(x,y)-P(x',y')$.

In our implementation, the Open GL projection technique runs faster than the ray casting projection technique (seconds vs tens of seconds). However, Open GL projection has a few problems. One of them is the resolution of the UV Map. Since the (u,v) coordinates are encoded in the color channel, there are only 256 possibilities for each coordinate. This could lead to discretization artifacts if the triangle spanned more than 256 pixels on the viewing

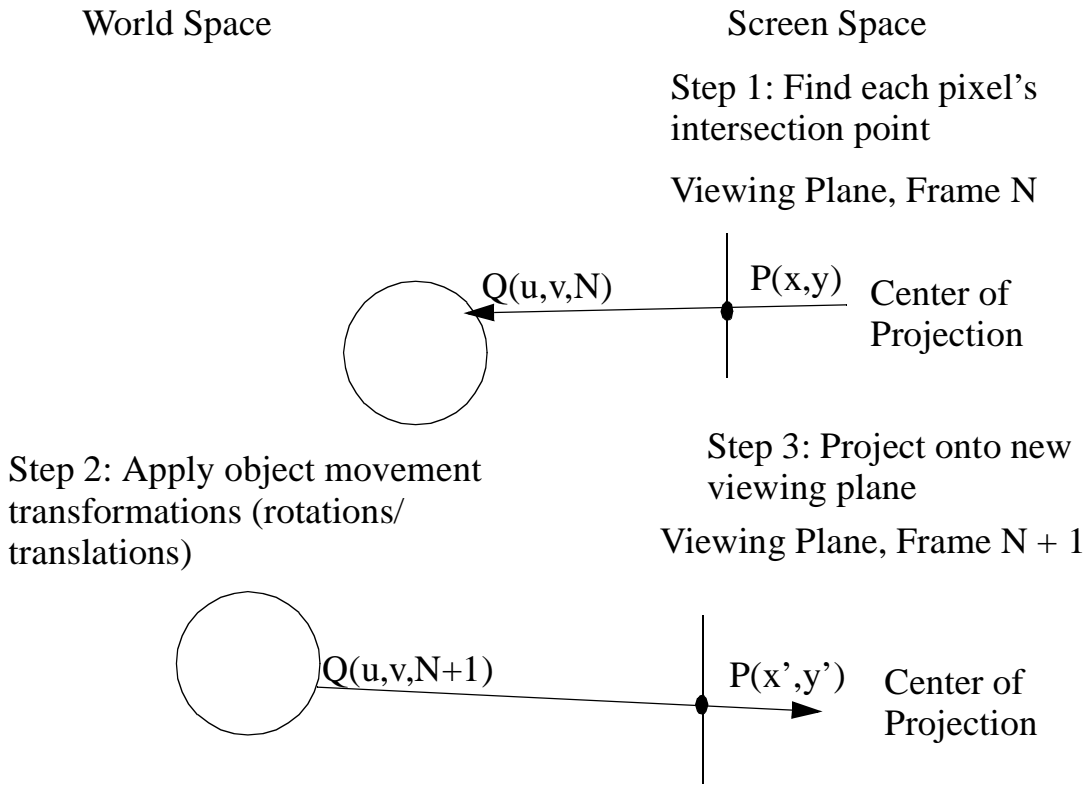

**Figure 5.6: Model-based Motion Estimation**

Each pixel P(x,y) on the viewing plane in Frame N is tracked to a corresponding object intersection location Q(u,v,N). The appropriate motion transformation is applied to derive Q(u,v,N+1) which is then projected onto the new viewing plane to obtain P(x',y'). From P(x,y) and P(x',y') we can calculate the pixel displacement from one frame to another.

plane. Discretization errors could also lead to incorrect displacement calculations.

For the sake of simplicity, the motion maps we use for computing spatiotemporal sensitivity will be derived from ray casting, model-based motion estimation. Figure 5.7 compares image-based and ray casting, model-based motion estimation techniques.

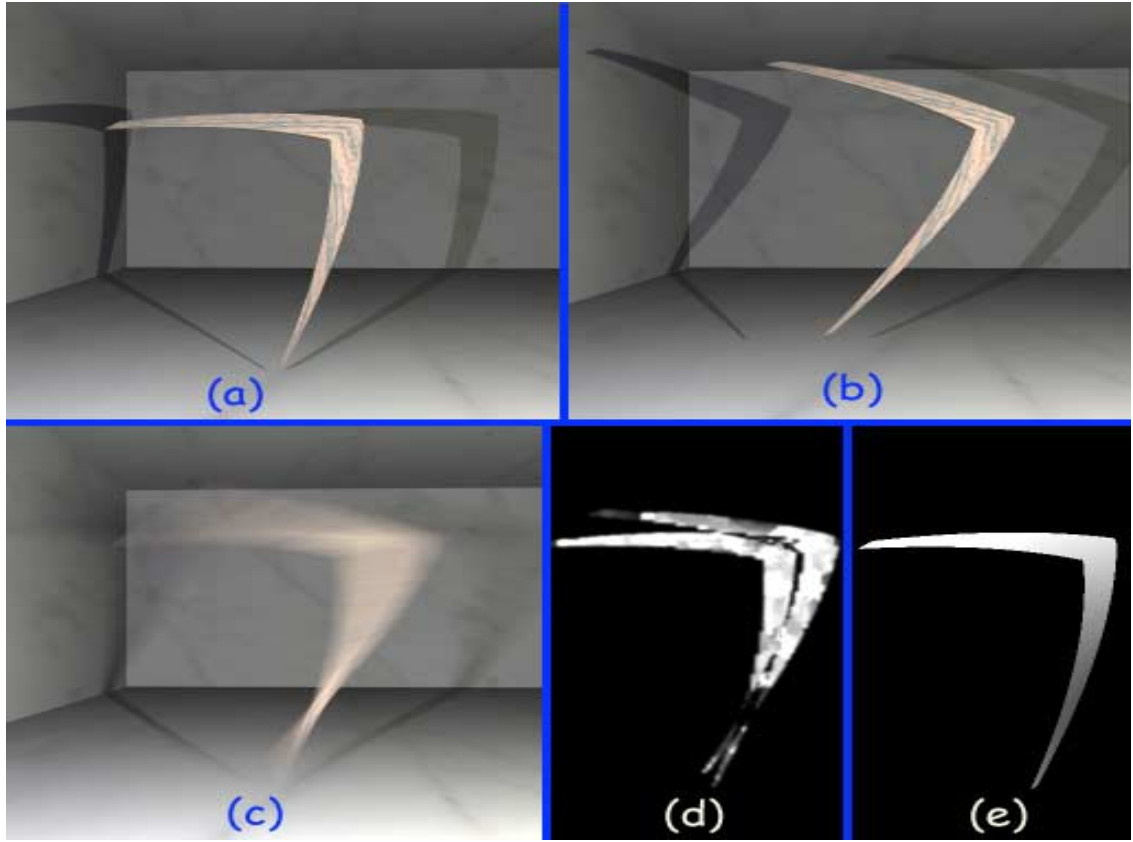

Figure 5.7: Comparison of Image-Based and Model-Based Motion Estimation

Two consecutive frames (a) and (b) are shown with the boomerang moving to the right from (a) to (b). The motion-blurred image in (c) shows the direction of motion. The results obtained using image-based motion estimation is shown in (d) and using ray casting, model-based motion estimation is shown in (e). For (d) and (e), the bright regions correspond to pixels of greater movement and the dark areas correspond to pixels that do not move. In (d), two copies of the boomerang appear because the image based motion estimation technique does not know if the boomerang is moving from left to right or if the disoccluded background is moving from right to left. Model-based motion estimation (e) is less noisy and more accurate than image-based motion estimation (d), which explains why (e) has a smooth motion estimation and (d) has a splotchy motion estimation.

### 5.1.3 Pixel Displacements and Motion

Using either image-based or model-based motion estimation, we have the pixel displacements $\Delta P(x,y)$ that tell us how much pixels move from one frame to another. However, the quantity we need for the sensitivity and saliency calculations is the image plane velocity. We convert the pixel displacements $\Delta P(x,y)$ computed by either of the two techniques into image plane velocities $v_I(x,y)$ using the following equation.

$$v_I(x, y) = \frac{\Delta P(x, y)}{\text{Pixels Per Degree}} \cdot \text{Frames per Second} \tag{5.1}$$

In our implementation, values were 30 frames per second on a display with a pixel density of 31 pixels per degree.

## 5.2 Spatial Frequency Estimation

Another component needed to calculate spatiotemporal error sensitivity is the spatial frequency content of the scene. The Fast Fourier Transform is usually used to obtain the frequency components of a signal but we have opted to use the faster Difference-of-Gaussians (Laplacian) Pyramid approach of Burt and Adelson [Burt83] to estimate spatial frequency content. One may reuse the Gaussian pyramid of the achromatic channel if it was computed in the image-based motion estimation step. Otherwise, the Gaussian pyramid is constructed by convolving the luminance channel of the image estimate with the separable filter {0.05,0.25,0.4,0.25,0.05}, downsampling by a factor of 2 and repeating the process as necessary. Each level of the Gaussian pyramid is upsampled to the size of the original image and then the absolute difference of the levels is computed to obtain the seven level bandpass Laplacian pyramid, L(0) to L(6).

$$L(i) = |A(i) - A(i+1)| \qquad (5.2)$$

The Laplacian pyramid has peak spatial frequency responses at $\rho_i = \{16, 8, 4, 2, 1, 0.5, 0.25\}$ cpd (assuming a pixel density of approximately 31 pixels per degree). Figure 5.8 graphically depicts one step of the Laplacian Pyramid calculation.

Using a method similar to that followed by Ramasubramanian, et al., [Rama99], each level of the Laplacian pyramid is then normalized by summing all the levels and dividing each level by the sum to obtain the estimation of the spatial frequency content in each frequency band:

$$R_i = \frac{L(i)}{\sum_{\text{all levels j}} L(j)} \qquad (5.3)$$

## 5.3 Saliency Estimation

The saliency estimation is executed using an extension of the computational model developed by Itti, et al., [Itti00][Itti98]. Our extension incorporates motion as an additional feature channel. The saliency map indicates locations of increased attention and is computed via the combination of four conspicuity maps of intensity, color, orientation and motion. The conspicuity maps are in turn computed using feature maps at varying spatial scales. One may think of features as stimuli at varying scales, conspicuity as a summary of a specific stimulus at all the scale levels combined and saliency as a summary of all the conspicuity of all the stimuli combined together. Figure 5.9

A(0) ⟶ [ Gaussian Filter ] ⟶ A(1)

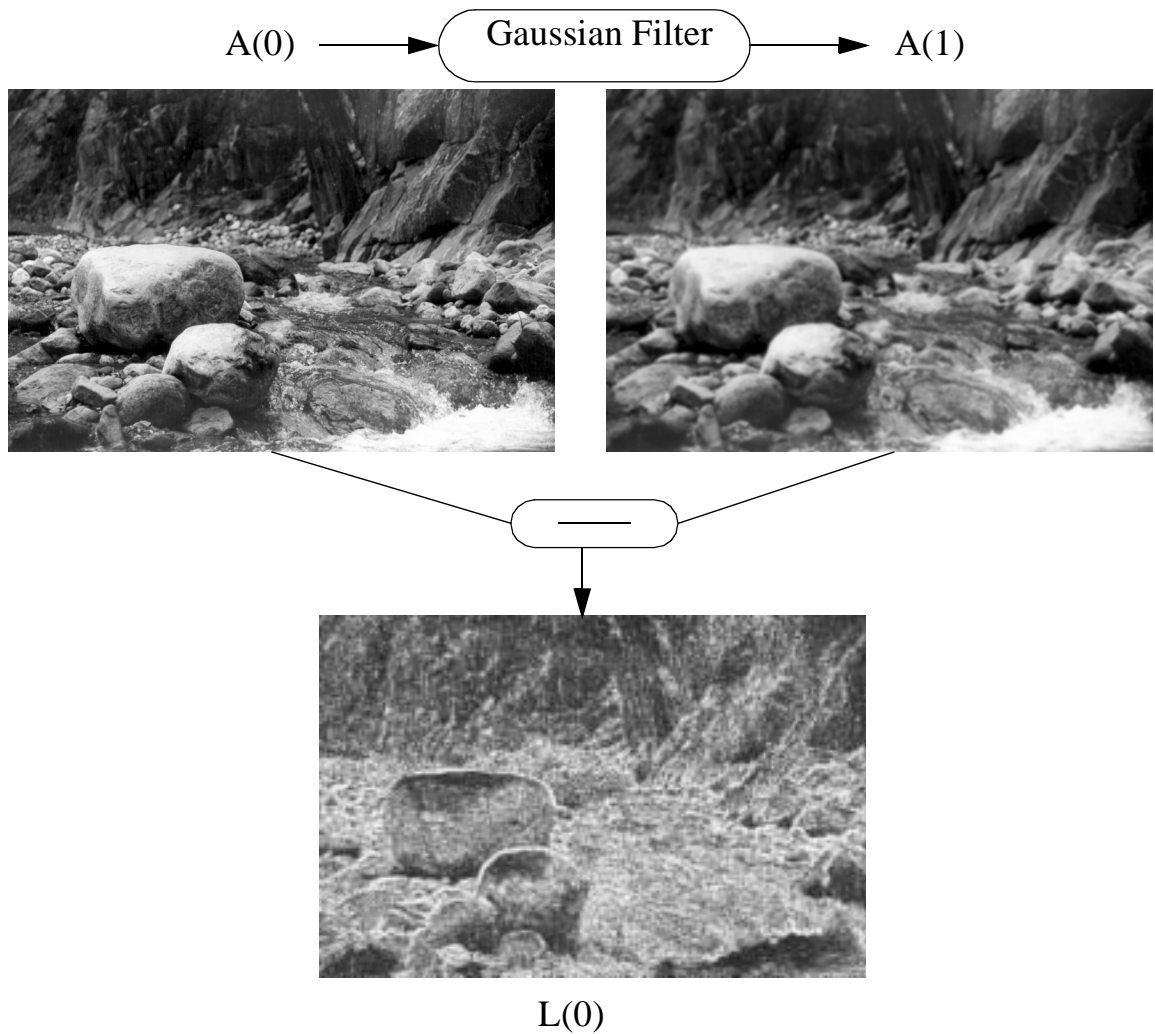

L(0)

Figure 5.8: Difference of Gaussians Operation

A single step of the Difference-of-Gaussians operation is shown. The achromatic luminance channel of the Image, A(0), is Gaussian filtered and then upsampled before an absolute difference is taken. The resulting image is the finest level of the Laplacian Pyramid, L(0).

documents the flow of computation from the image estimate to the saliency map.

Feature maps for the achromatic (A) and chromatic ($C_1$,$C_2$) channels are computed by constructing image pyramids similar to the Laplacian pyramid described in the previous section. A Gaussian pyramid is constructed for each

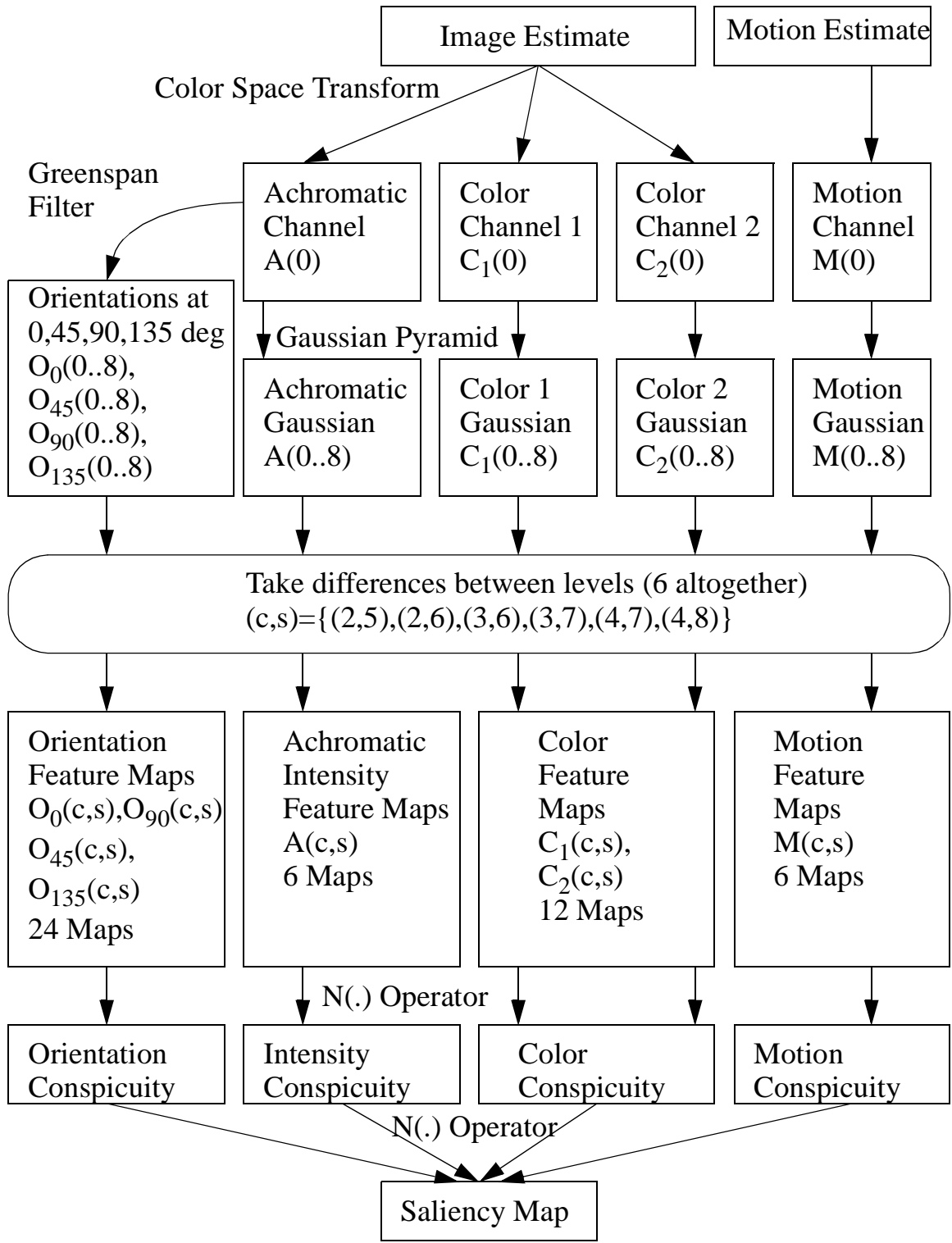

Figure 5.9: Saliency Map Computation

A flowchart of the Saliency Map computation process.

channel and following Itti, et al., we obtain the feature maps in the following manner:

$$X(\text{center, surround}) \; = \; |X(\text{center}) - X(\text{surround})| \qquad (5.4)$$

where X stands for $A, C_1, C_2$ and (center,surround) $\in$ {(2,5), (2,6), (3,6), (3,7), (4,7), (4,8)}. The numbers correspond to the levels in the Laplacian pyramid.

Motion feature maps are created by applying a similar decomposition to the velocity map generated in the motion estimation section.

Orientation feature maps are obtained by creating four pyramids using Greenspan's [Gree94] filter on the achromatic channel. Greenspan's filter was tuned to orientations of (0, 45, 90 and 135 degrees) and indicates what components of the image lie along those orientations. We generate a total of 48 feature maps, 6 for intensity at different spatial scales, 12 for color, 6 for motion, and 24 for orientation for determining the saliency map. The feature maps are a multiple of six due to the number of combinations of center-surround operations in Equation 5.4.

Next, we combine these feature maps to get the conspicuity maps and then combine the conspicuity maps to obtain a single saliency map for each image frame. We use a global non-linear normalization operator, N(.), described in [Itti98] to simulate lateral inhibition and then sum the maps together to perform this combination. This operator carries out the following operations:

1. Normalize each map to the same dynamic range, e.g. (0..1).

2. Find the global maximum M and the average $\overline{m}$ of all other local maxima. Local maxima are defined as pixels whose values are greater than pixels in its immediate neighborhood.

3. Scale the entire map by $(M-\overline{m})^2$.

The purpose of the N(.) operator is to promote maps with significantly conspicuous features while suppressing those that are non-conspicuous. Figure 5.10 illustrates the action of the N(.) operator on three generic maps.

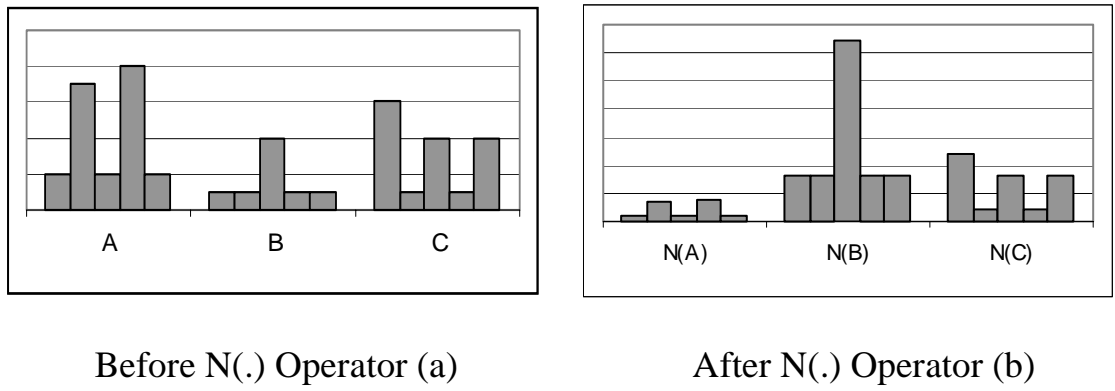

Before N(.) Operator (a)                    After N(.) Operator (b)

Figure 5.10: Action of the N(.) Operator

The left half (a) shows the maps after step 1. The right half (b) shows the maps after steps 2 and 3. Map A and C have competing signals and are suppressed. Map B has a clear spike and is therefore promoted. In this way, the N(.) operator roughly simulates the lateral inhibition behavior of the visual system. When N(.) is applied to feature maps, A,B,C represent the levels of the corresponding Laplacian pyramid of the feature. When applied to conspicuity maps, A,B and C represent channels such as intensity or color.

We apply the N(.) operator to each feature map and combine the resulting maps of each channel's pyramid into a conspicuity map. We now get the four conspicuity maps of intensity, color, orientation and motion. We then compute the saliency map by applying N(.) to each of the four conspicuity maps and then summing them together. We will call the saliency map S(x,y) with the per pixel saliency normalized to a range of (0.0... 1.0) where one represents the most salient region and zero represents the least salient region in the image. Figure 5.11 shows the saliency map computed for one of the animation image frames.

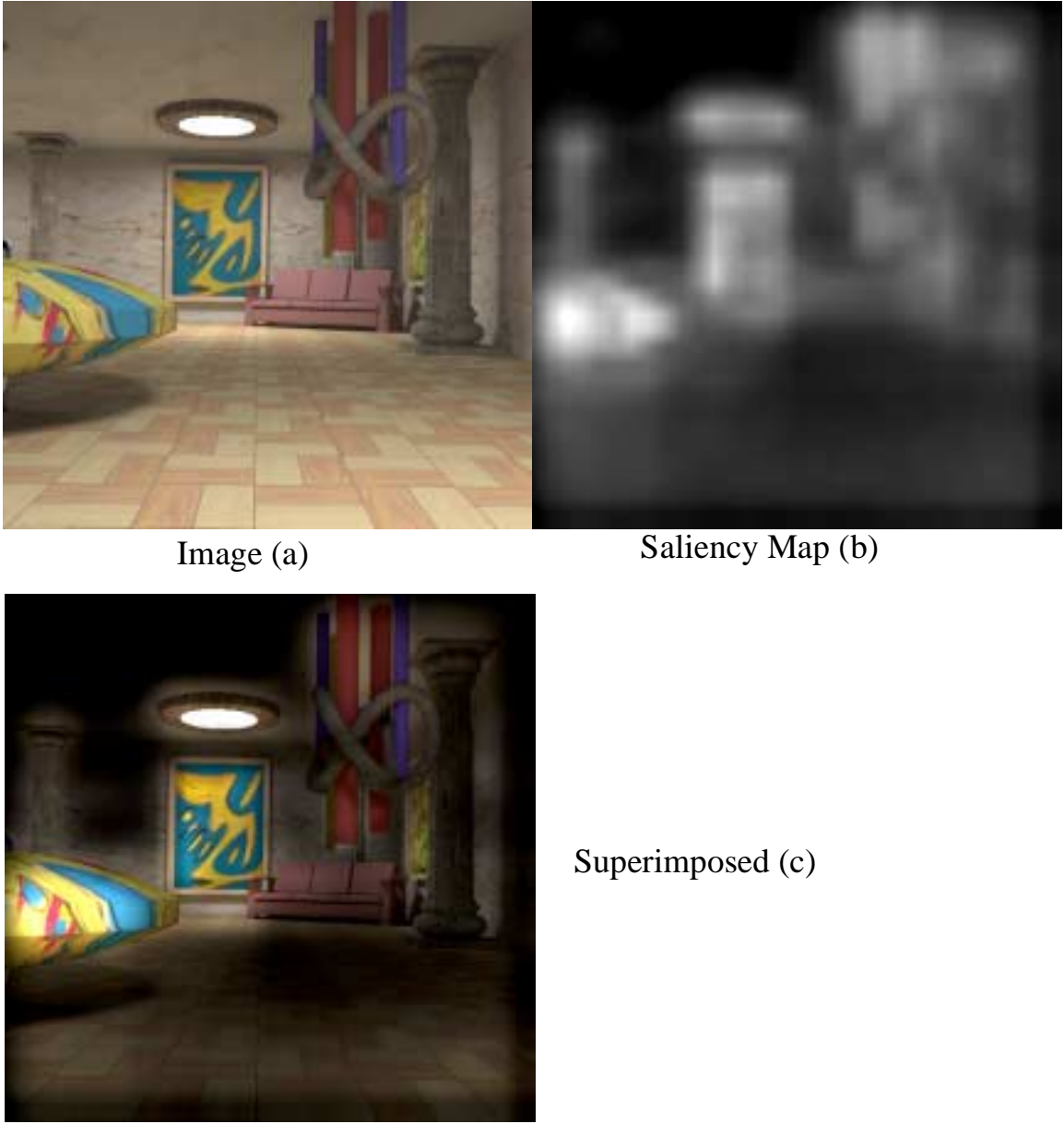

Image (a)                        Saliency Map (b)

Superimposed (c)

Figure 5.11: Saliency Map Visualization

In image (a) the yellow and blue top on the left is spinning rapidly. The computed saliency map is shown in (b) and (c) graphically depicts the modulation of the saliency map with the image. Brighter areas denote areas of greater saliency. Attention is drawn strongly to the spinning top, the paintings, the ceiling sculpture, the area light and the couch. These areas undergo strict motion compensation. The floor and ceiling are not as salient and undergo less compensation.

## 5.4 Spatiotemporal Error Sensitivity Computation

At this stage, we will have the weights for spatial frequency from the bandpass responses $R_i(x,y)$ (Equation 5.4) with peak frequencies $\rho_i = \{16,8,4,2,1,0.5,0.25\}$ cycles per degree, the image plane pixel velocities $v_I(x,y)$ (Equation 5.1), and the saliency map $S(x,y)$. We now have all the necessary ingredients to estimate the spatiotemporal sensitivity of the HVS. The first step is to obtain the retinal velocity $v_R$ from the image plane velocity $v_I$ with the use of the saliency map $S(x,y)$ to modulate image plane velocity:

$$v_R(x, y) = v_I(x, y) - min(S(x, y) \cdot v_I(x, y) + v_{Min}, v_{Max}) \qquad (5.5)$$

where $v_{Min}$ is the drift velocity of the eye (0.15 deg/sec [Kell79]) and $v_{Max}$ is the maximum velocity beyond which the eye cannot track moving objects efficiently (80 deg/sec [Daly98]). We use this velocity to compute the spatiotemporal sensitivities at each of the spatial frequency bands $\rho_i$. For this computation, we use Kelly's experimentally derived contrast sensitivity function (CSF):

$$CSF(\rho, v_R) = k \cdot c_0 \cdot c_2 \cdot v_R \cdot (2\pi\rho c_1)^2 \cdot e^{-(4\pi c_1 \rho)/\rho_{max}} \qquad (5.6)$$

$$k = 6.1 + 7.3 \left| \log((c_2 \cdot v_R)/3) \right|^3 \qquad (5.7)$$

$$\rho_{max} = (45.9)/(c_2 \cdot v_R + 2) \qquad (5.8)$$

Following the suggestions of Daly [Daly98], we set $c_0$=1.14, $c_1$=0.67 and $c_2$=1.7. These parameters are tuned to CRT display luminance of 100 cd/m$^2$.

Contrast sensitivity is the inverse of threshold contrast. Therefore, the inverse of the CSF intuitively gives us an elevation factor that increases our tolerance of error beyond the minimum discernible luminance threshold in optimal viewing conditions. We calculate this elevation factor for each of the peak spatial frequencies of our Laplacian pyramid $\rho_i \in \{16,8,4,2,1,0.5,0.25\}$ cpd:

$$f_i(\rho_i, v_R) = \begin{array}{l} \dfrac{CSF_{Max}(v_R)}{CSF(\rho_i, v_R)} \quad \text{if } (\rho_i > \rho_{max}) \\[2mm] 1.0 \qquad\qquad \text{otherwise} \end{array} \qquad (5.9)$$

$$CSF_{Max}(v_R) = \frac{\rho_{Max}}{2\pi c_1} \qquad (5.10)$$

where $v_R$ is the retinal velocity, CSF is the spatiotemporal sensitivity function, $CSF_{Max}(v_R)$ is the maximum value of the CSF at velocity $v_R$, and $\rho_{max}$ is the spatial frequency at which this maximum occurs.

Finally we compute the Aleph Map, the spatiotemporal error tolerance map, as a weighted sum of the elevation factors $f_i$, and the frequency responses $R_i$ at each location (x,y):

$$\aleph(x, y) = \sum_i R_i \times f_i \qquad (5.11)$$

The computation of Equations 5.9 - 5.11 are similar to the computation of the threshold elevation map described in [Rama99] with the difference that the CSF function used here is the spatiotemporal CSF instead of the spatial only CSF. Figure 5.12 shows the error tolerance map $\aleph$(x,y) for an image frame of a dynamic scene. This map captures the sensitivity of the HVS to the spatiotemporal contents of a scene. $\aleph$(x,y) has values ranging from 1.0 (lowest tolerance to error) to at most 250.0 (most tolerance to error). The value $\aleph$(x,y) represents the contrast elevation factor due to spatial and temporal frequencies that increases the contrast needed to discern a signal from the background.

The next chapter will demonstrate how the Aleph Map can be adapted for use in accelerating global illumination.

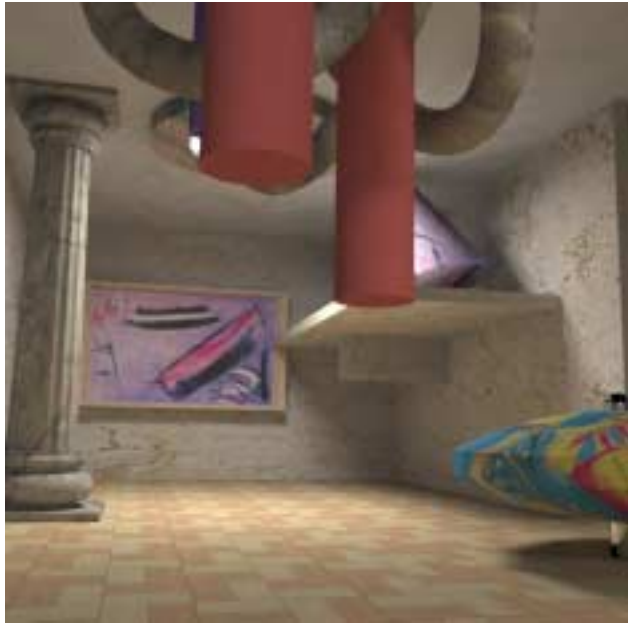

Image (a)

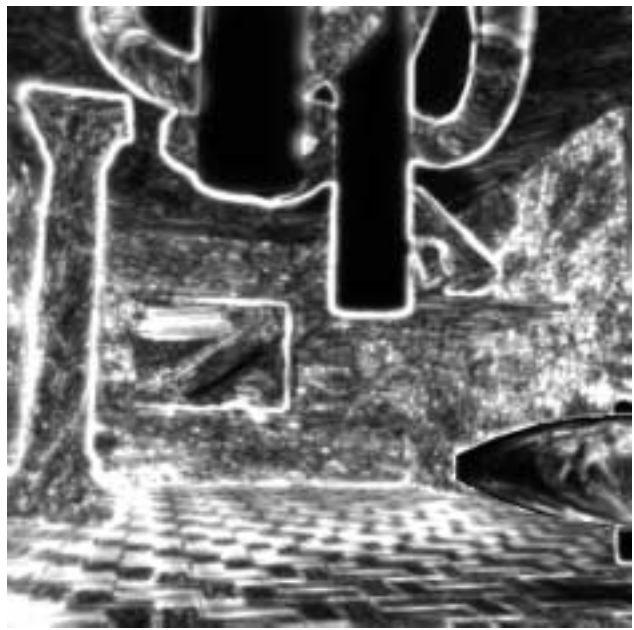

Aleph Map (b)

Figure 5.12: Spatiotemporal Error Sensitivity Visualization

Image (a) and its corresponding error tolerance map, the Aleph Map (b). Note that the spinning top in the bottom right has reduced tolerance to error although it has textures and is moving. This is due to the information introduced by the saliency map, telling the algorithm to be stricter on the top because the viewer will more likely focus attention there. The red beams are treated strictly because there are no high frequency details.

# CHAPTER 6
# Applications & Results

*"If you want to make an apple pie from scratch, you must first create the universe." - Carl Sagan*

## 6.1 Application to Irradiance Caching

The Aleph Map developed in the previous sections is general. It operates on image estimates of any animation sequence to predict the relative error tolerance at every location of the image frame and can be used to efficiently render dynamic environments. Similar to earlier perceptually-based acceleration techniques [Boli95][Boli98][Mysz98][Rama99], we can use this map to adaptively stop computation in a progressive global illumination algorithm. On the other hand, we can also use the map as a perceptual oracle to specify in advance the amount of computation to apply to a lighting problem. To demonstrate the wider usefulness of this map we have applied the map to improve the computational efficiency of irradiance caching, the key algorithm behind the widely used program RADIANCE.

The irradiance caching algorithm is the core technique used by RADIANCE to accelerate global illumination and is well documented by Ward [Ward88][Ward92][Ward98]. As suggested by its name, the irradiance caching technique works by caching the diffuse indirect illumination component of global illumination [Ward88]. A global illumination lighting solution can be calculated as the sum of a direct illumination term and an indirect illumination term. Indirect illumination is by far the most computationally expensive portion of the calculation, and is usually computed using Monte Carlo evaluation of thousands of light samples. Irradiance caching addresses this problem by reusing irradiance values from nearby locations in object space and interpolating them, provided the error that results from doing so is bounded by the evaluation of an ambient accuracy term. Hence, by reusing information, the irradiance caching algorithm is faster than the standard Monte Carlo simulation of the global illumination problem by several orders of magnitude, while at the same time providing a solution that has bounded error.

Figure 6.1 diagrams the operation of the irradiance cache. In the figure, E1 and E2 are irradiance values that were calculated previously and stored in the cache. Each irradiance value has a radius within which it is valid. This radius is determined from the ambient accuracy term and the harmonic distance of the location of the irradiance value from the other surfaces around it. When a nearby irradiance value is needed, the cache is checked to determine if there are irradiances in it that can be used to interpolate the new irradiance. In the figure, test points A and B lie within one or more stored irradiance values and can be calculated from interpolation. On the other hand, test point C lies outside of the valid radius of any irradiance value and must be calculated by Monte Carlo evaluation. After being computed, the irradiance at test point C is stored in the irradiance cache for future use.

A new irradiance value E at a point P can be computed from cached irradiance values in the following manner [Ward98]:

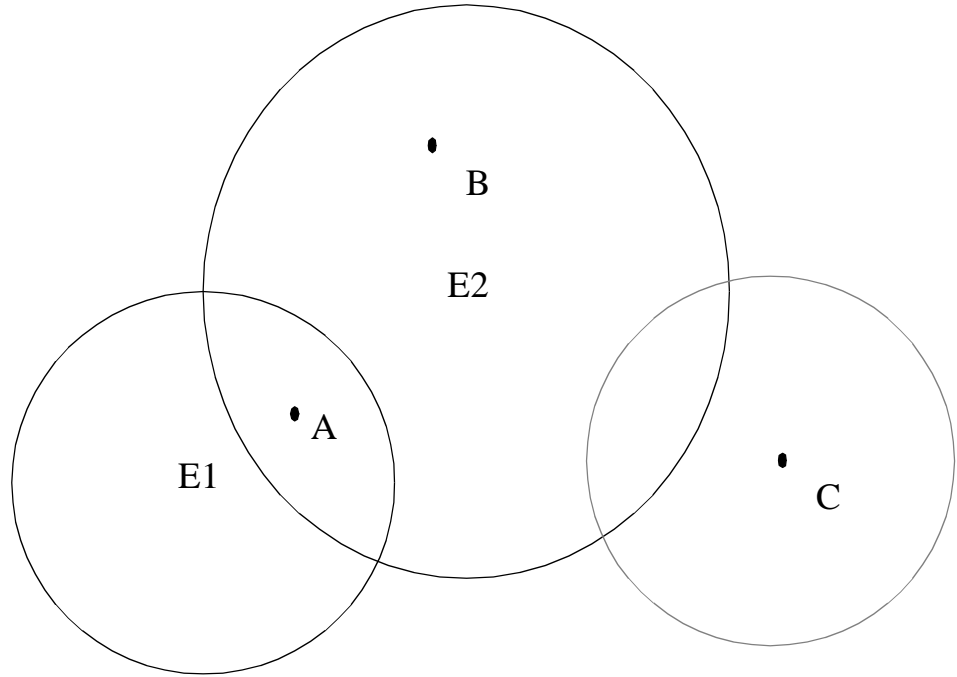

Figure 6.1: Irradiance Cache

Indirect irradiances E1 and E2 were calculated previously and stored in the irradiance cache, together with their validity radii. Test points A and B are close enough to existing cache values to be computed via extrapolation (Equation 6.1), but point C requires the calculation of a new irradiance value. Modified from [Ward98].

$$E(\vec{P}) \; = \; \frac{\displaystyle\sum_{i \in S} w_i(\vec{P}) E_i(\vec{P})}{\displaystyle\sum_{i \in S} w_i(\vec{P})} \tag{6.1}$$

where

$$w_i(\vec{P}) \; = \; \left( \frac{\left\| \vec{P} - \vec{P_i} \right\|}{R_i} + \sqrt{1 - \vec{N}(\vec{P}) \bullet \vec{N}(\vec{P_i})} \right)^{-1} \tag{6.2}$$

$$\vec{N}(\vec{P}) \; = \; \text{surface normal at position P} \tag{6.3}$$

$$E_i(\vec{P}) = \text{computed illuminance at P}_i \text{ extrapolated to P} \qquad (6.4)$$

$$R_i = \text{harmonic mean distance to objects visible from P}_i \qquad (6.5)$$

$$S = \langle i | w_i(\vec{P}) > \frac{1}{\alpha_{\text{Acc}}} \rangle \qquad (6.6)$$

Equation 6.2 is the inverse of the error term derived from modeling the error on a sphere that is lit on one side and dark on the other. This 'split sphere' error models the worst error that can occur when using irradiance interpolation. The inverse of the error term is used to weight the irradiance cache values in order to derive the extrapolated irradiance value. Note that the error is a function of the distance of the cache value from the test location as well as the orientation of the cache value with respect to the surface normal at the test location. The ambient accuracy term, $\alpha_{\text{Acc}}$, is user supplied and provides a control on the error allowed for indirect illumination by picking the valid irradiance cache values for extrapolation in Equation 6.6. The ambient accuracy term varies from 0.0 (no interpolation, purely Monte Carlo simulation) to 1.0 (maximum ambient error allowed). The as $\alpha_{\text{Acc}}$ get smaller, the set S reduces in the number of elements it contains until it becomes an empty set, whereupon Monte Carlo evaluation is used to compute the irradiance value. In the implementation of irradiance caching, $\alpha_{\text{Acc}}$ is also used to modulate the domain of influence of the irradiance value. Smaller values of $\alpha_{\text{Acc}}$ reduce the radius over which the cached irradiance value is valid.

RADIANCE uses the ambient accuracy term uniformly over the entire image, and thus does not take advantage of the variation of sensitivity of the HVS over different parts of the image. Our application of the Aleph Map to the irradiance caching algorithm works by modulating the ambient accuracy term on a per pixel basis. Hence, wherever the Aleph Map allows for greater error for that pixel, a larger set of irradiance values are considered for interpolation, making efficient use of the irradiance cache. In order to use the Aleph Map with the irradiance cache we need to use a compression function to map the values of $\aleph(x,y)$ onto $(\alpha_{Acc} -1.0)$ for use as a perceptual ambient accuracy term. The following equation accomplishes this compression:

$$\alpha_1(x, y) \;=\; \frac{\aleph(x, y)}{\aleph(x, y) - 1 + \dfrac{1}{\alpha_{Acc}}} \tag{6.7}$$

where $\alpha_1$ is the adapted map used in lieu of the original ambient accuracy term $\alpha_{Acc}$. Figure 6.2 plots a graph of the compression function. The equation is a heuristic that ensures $\alpha_1$ is bounded between $\alpha_{Acc}$ and 1.0. Hence, in regions where attention is focused and where there are no high frequencies to mask errors, $\alpha_1 = \alpha_{Acc}$ and in areas where the errors will be masked, $\alpha_1$ asymptotically approaches 1.0. Computation of $\alpha_1$ is carried out only once, at the beginning of the global illumination computation of every frame. However, should a stricter bound be desired, one may opt to recompute $\aleph(x,y)$ and hence recompute $\alpha_1$ at intermediate stages of computation.

A simpler heuristic can also be used to convert $\aleph(x,y)$ into a form usable by irradiance caching by simple scaling and adding:

$$\alpha_2 \;=\; \alpha_{Acc} + \frac{\aleph}{K} \tag{6.8}$$

where K is some suitable constant that scales $\aleph(x,y)$. Since the maximum value of the CSF is about 250, one may chose K=250, or in our case, K=100. Note that using this scaling and adding heuristic, there is no guarantee that $\alpha_2$ is bounded by 1.0 from above.

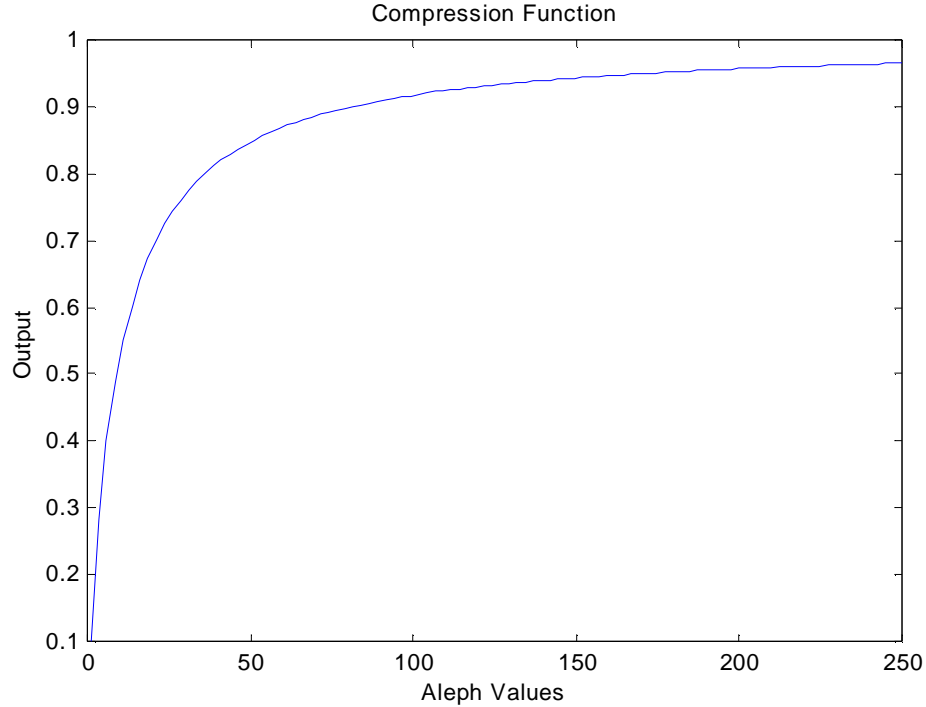

**Figure 6.2: Compression Function**

The compression function (Equation 6.7) maps the contrast values of the Aleph Map onto the range of ($\alpha_{Acc}$ to 1.0) for use as a perceptually-based ambient accuracy term.

A dynamic simulation of a pool ball collision was used to select a suitable heuristic to convert Aleph Map values into ambient accuracy values. Figure 6.3 shows the performance speedups on a the pool table sequence rendered at 8192 samples per irradiance value with a base ambient accuracy of $\alpha_{Acc}$=0.1.

In the figure, full refers to the full motion compensation heuristic in Equation 3.1 (page 22). Saliency refers to the saliency map based motion compensation heuristic in Equation 5.5 (page 50). In either case, the appropriately compensated image plane velocity is used to derive the Aleph Map and an adapter function is used to convert Aleph Map values into a perceptually-based ambient accuracy term. The compressive function (Equation 6.7) pro-

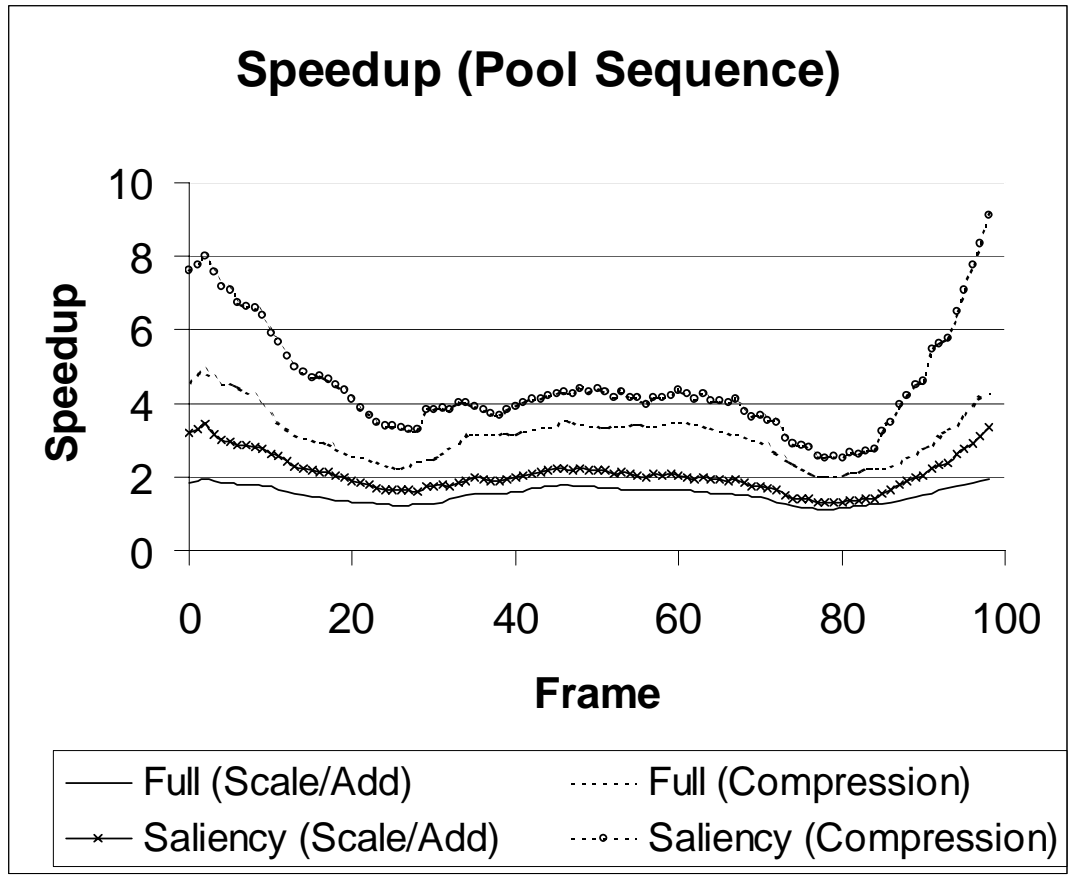

Figure 6.3: Pool Sequence Performance Data

The speedups are relative to a reference solution rendered with 8192 samples per irradiance value with a base ambient accuracy of 0.1. Full refers to the full motion compensation heuristic which assumes that the eye tracks everything equally well (Equation 3.1, page 22). Saliency refers to using the saliency map to determine the eye's tracking velocity (Equation 5.5, page 50). Compression is the heuristic in Equation 6.7, and Scale/Add is the heuristic in Equation 6.8.

vides a better speedup in irradiance caching over the scale/add function (Equation 6.8) when used to transform Aleph Map values to a perceptually based ambient accuracy term. In subsequent uses of the Aleph Map in irradiance caching, the compressive function in Equation 6.7 will be used in favor of the scale/add function for calculating the perceptually based ambient accu-racy term. Figure 6.4 demonstrates the visual performance of the compressive function on the pool table sequence.

We further demonstrate the performance of our enhancement using a test scene of a synthetic art gallery. The scene contains approximately 70,000 primitives and eight area light sources. It contains many moving objects, including bouncing balls, a spinning top and a kinetic sculpture that demonstrates color bleeding on a moving object. Figures 6.5 to 6.9 show a visual comparison of a reference solution and it's corresponding Aleph Map accelerated solution. In the figures, the reference solution was rendered with 8192 samples per irradiance value and a base ambient accuracy of 0.15. The Aleph Map accelerated solution was rendered using Saliency based motion compensation and uses the compression function (Equation 6.7) to map Aleph Map values onto a perceptually-based ambient accuracy term. Speedups of an order of a magnitude were achieved. The reference solution takes between four to six hours per frame to compute while the Aleph Map solution takes between 20 minutes to an hour to compute. Times are for a single 550 MHZ Pentium III quad processor node.

Figure 6.10 shows the performance improvement resulting from the use of the Aleph Map on the Art Gallery sequence. The figure compares the performance, as measured in sampling efficiency, compared to irradiance caching. Spatial factors only indicate that the scene was rendered with an Aleph Map using image plane velocities that were set to zero. Full motion compensation indicates that the scene was rendered with the Aleph Map's velocity component motion compensated using Daly's equation (Equation 3.1, page 22). Aleph Map indicates that the scene was rendered using an Aleph Map with image plane velocities compensated using the Saliency map. The spatial only, full compensation and Aleph Map images were rendered at a resolution of

Reference       Full
Compensation       Saliency Map
Compensation

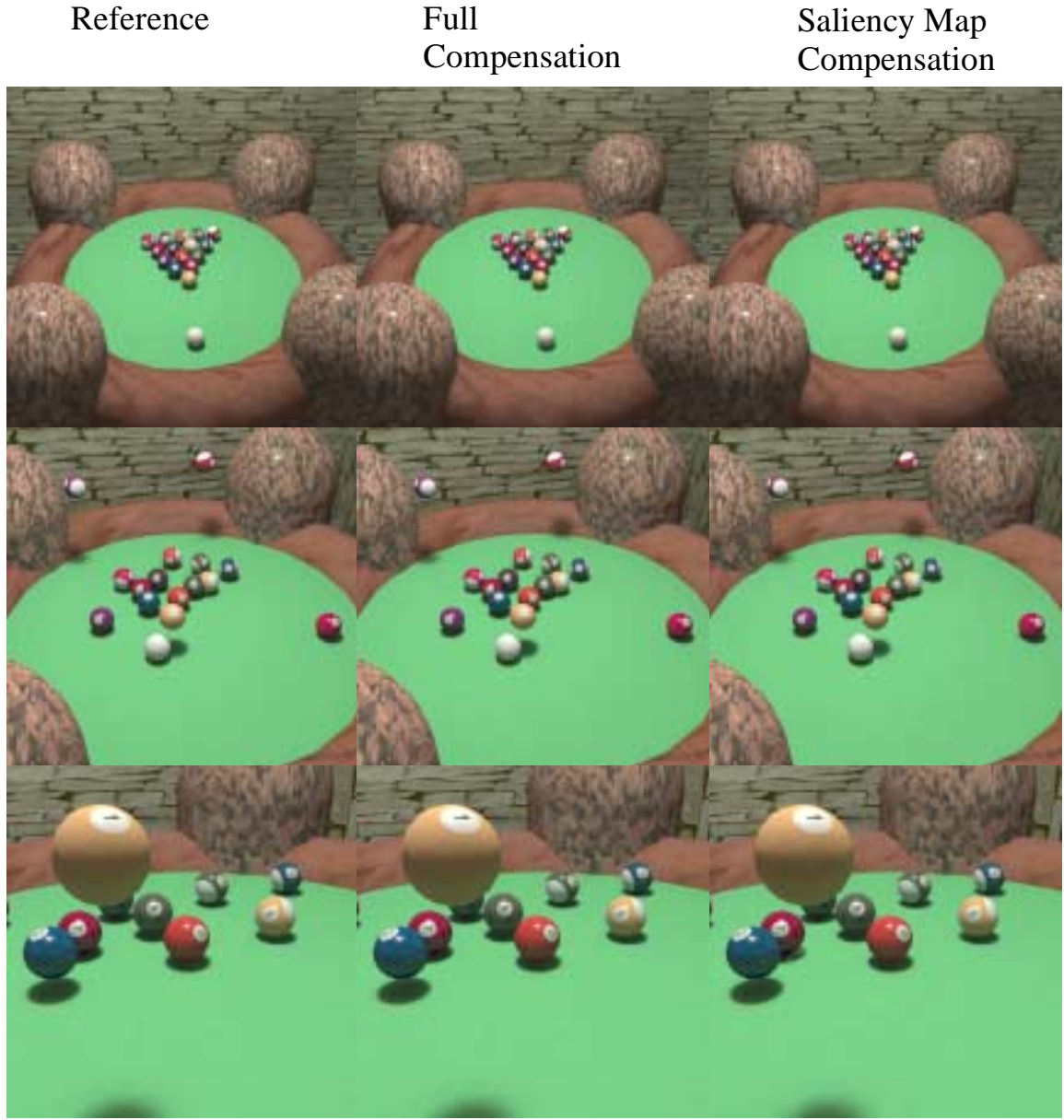

Figure 6.4: Pool Sequence Visual Comparison

The left column displays a few frames of the reference solution from the Pool
sequence. The middle column shows images derived using full motion
compensation and the compressive function (Equation 6.7) that maps the
Aleph map to a per-pixel perceptually-based ambient accuracy term. The
right column shows images derived using Saliency-based motion
compensation and the same compressive function (Equation 6.7).

Reference

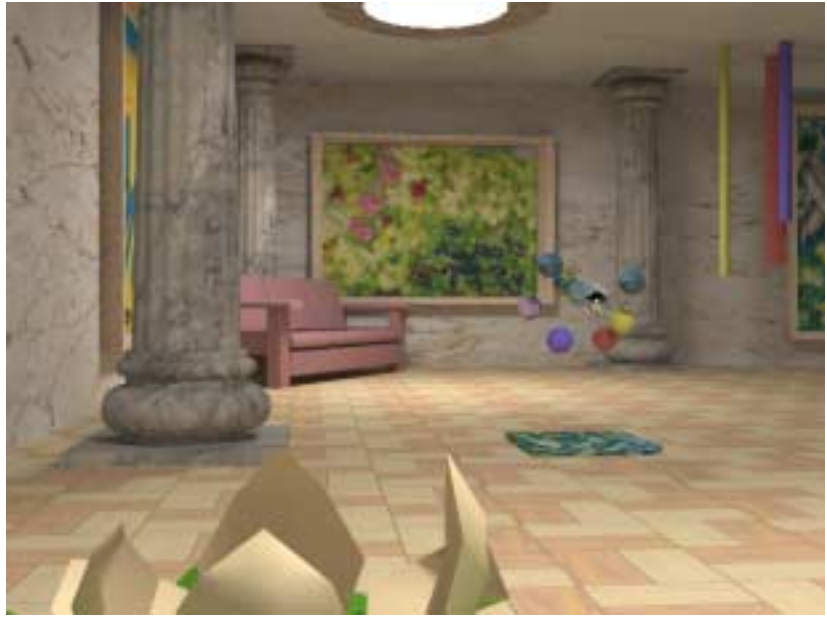

Aleph Map Accelerated Solution

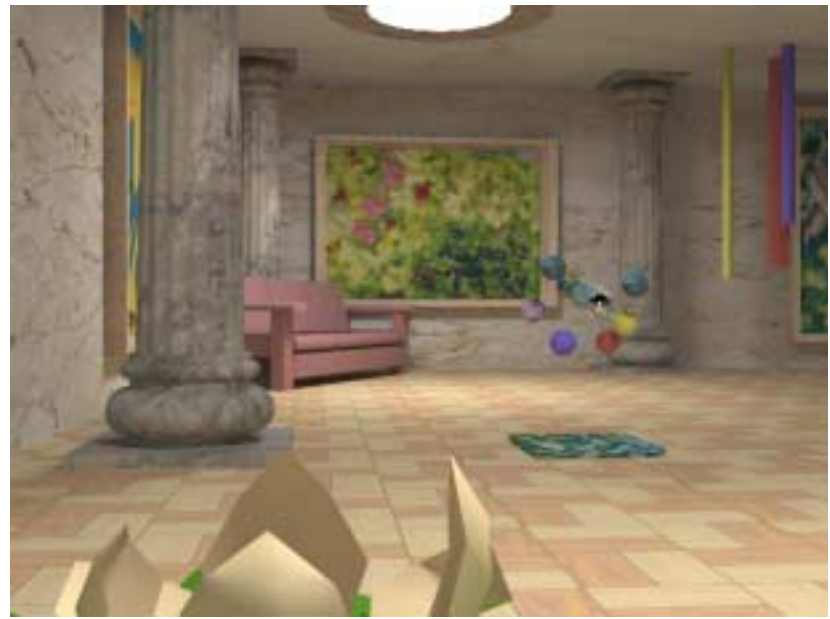

Figure 6.5: Art Gallery, Frame 0.

Comparison between a reference solution (above) and the Aleph Map accelerated version (below). The Aleph Map accelerated irradiance cache performs seven times better than the reference solution in this frame.

Reference

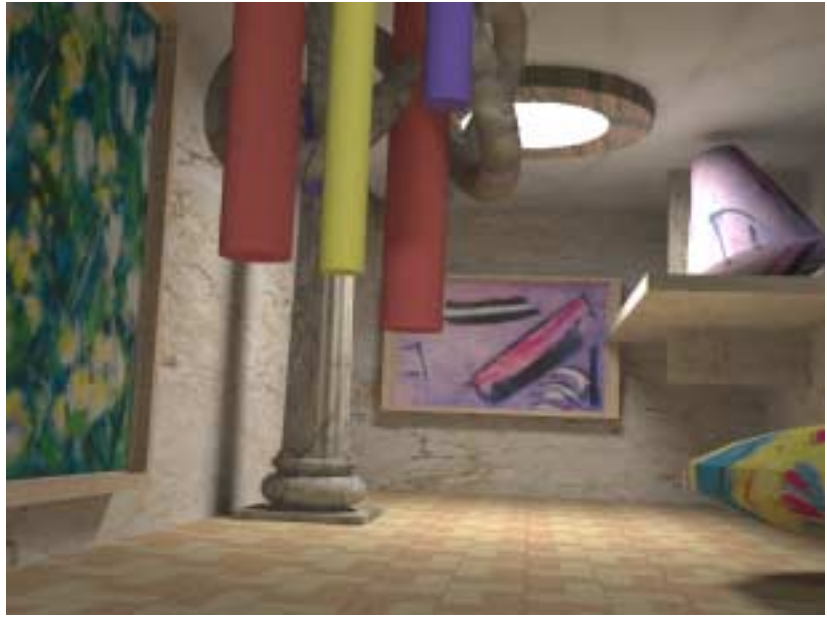

Aleph Map Accelerated Solution

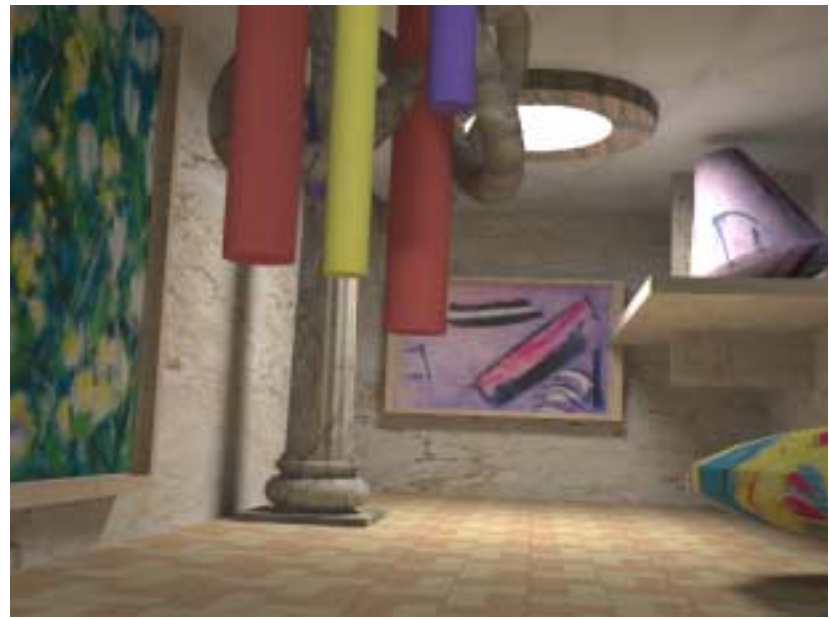

Figure 6.6: Art Gallery, Frame 180.

Comparison between a reference solution (above) and the Aleph Map accelerated version (below). The Aleph Map accelerated irradiance cache performs eight times better than the reference solution in this frame.

Reference

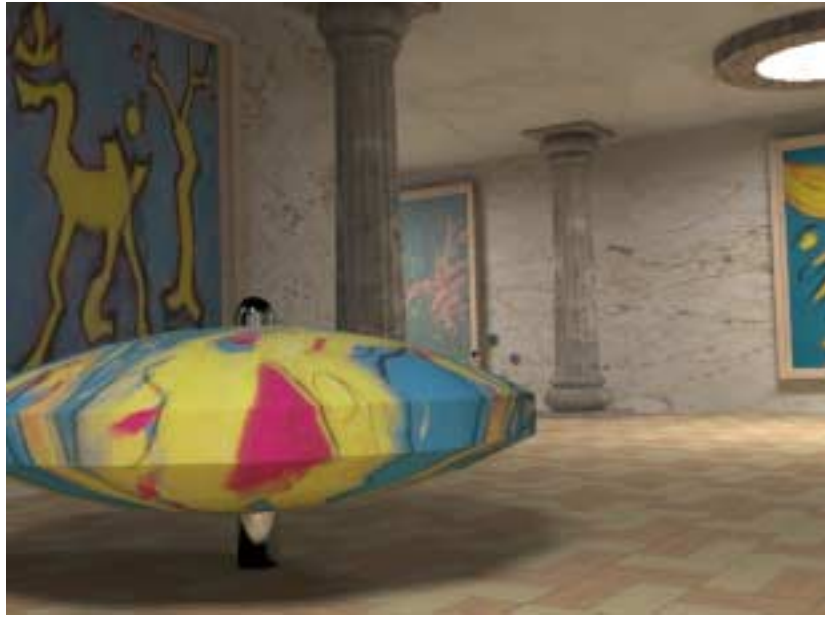

Aleph Map Accelerated Solution

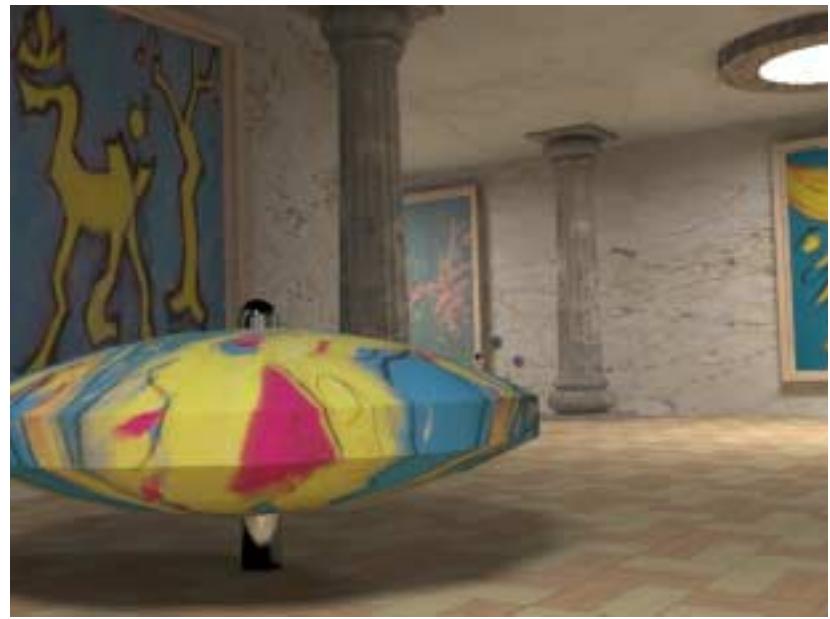

Figure 6.7: Art Gallery, Frame 300.

Comparison between a reference solution (above) and the Aleph Map accelerated version (below). The Aleph Map accelerated irradiance cache performs 11 times better than the reference solution in this frame.

Reference

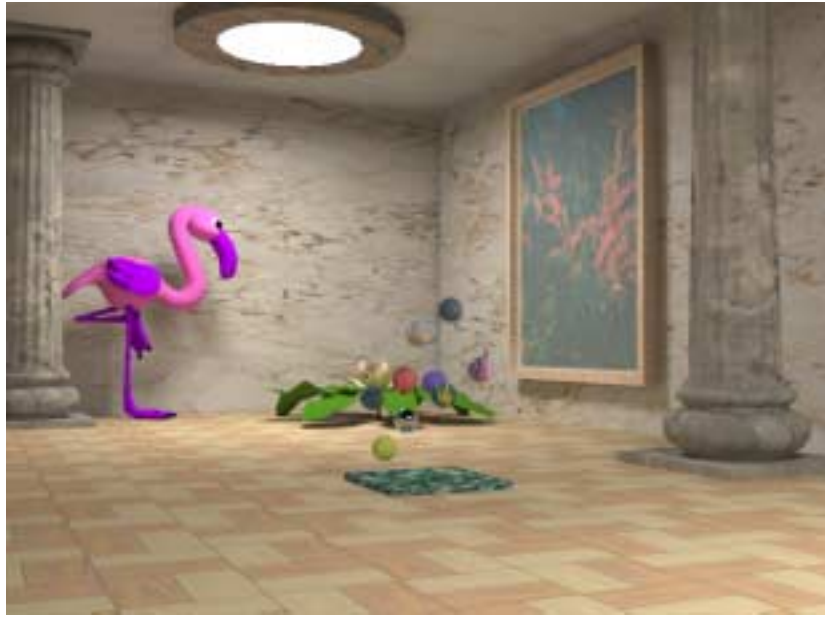

Aleph Map Accelerated Solution

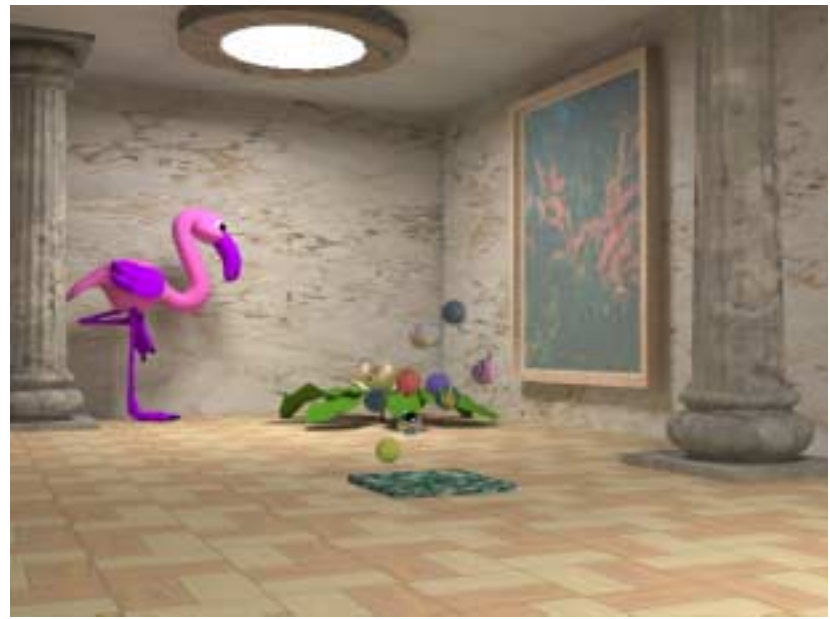

Figure 6.8: Art Gallery, Frame 540.

Comparison between a reference solution (above) and the Aleph Map accelerated version (below). The Aleph Map accelerated irradiance cache performs nine times better than the reference solution in this frame.

Reference

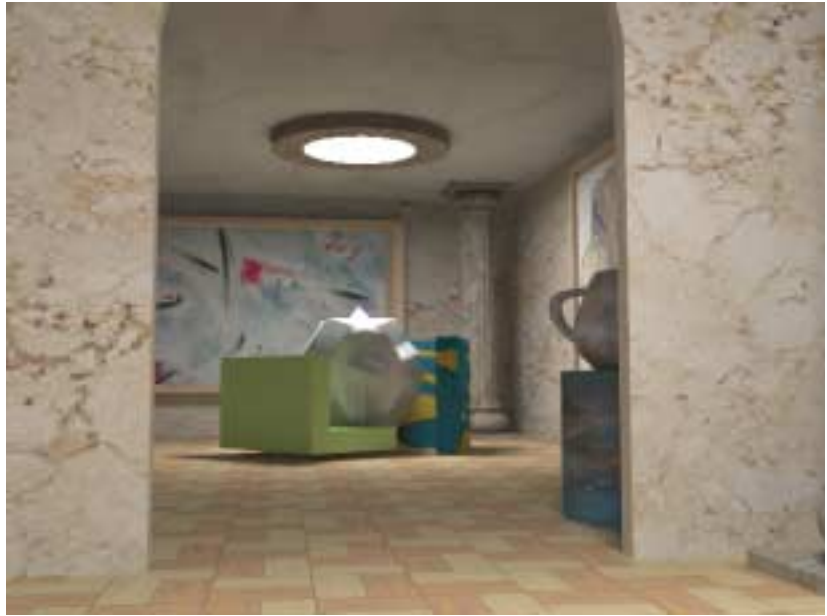

Aleph Map Accelerated Solution

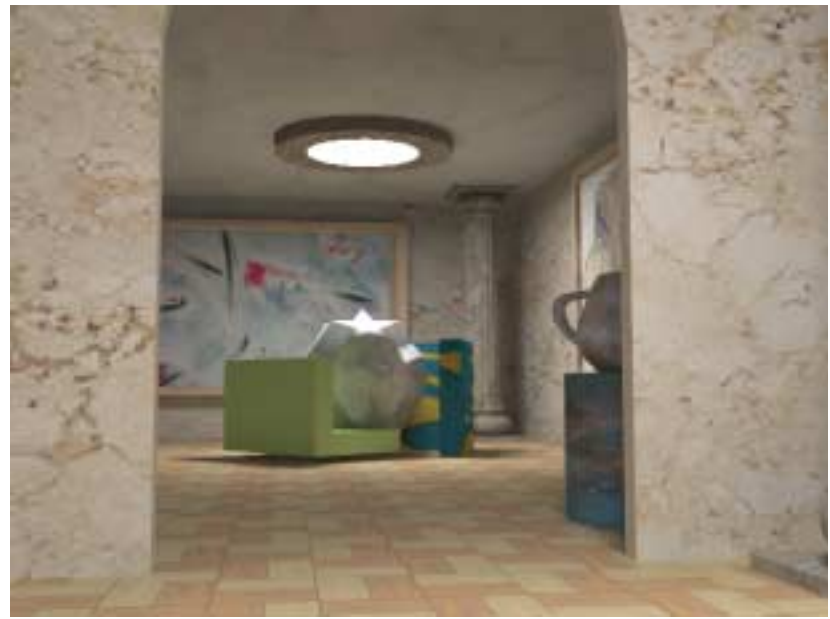

Figure 6.9: Art Gallery, Frame 720.

Comparison between a reference solution (above) and the Aleph Map accelerated version (below). The Aleph Map accelerated irradiance cache performs seven times better than the reference solution in this frame.

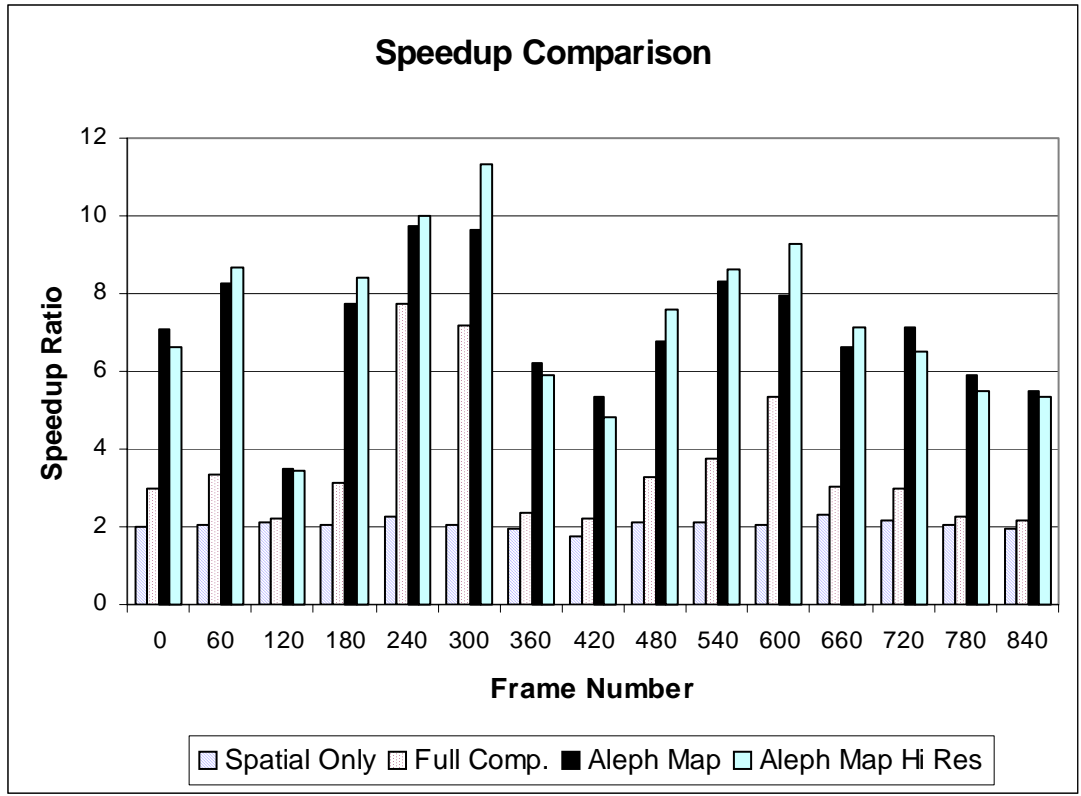

Figure 6.10: Speedup over Irradiance Cache for the Art Gallery sequence.

The total number of ray triangle intersections per pixel are compared. The Aleph Map enhanced irradiance cache performs significantly better (6-8x) than the unaugmented irradiance cache. Spatial factors contribute to an average of 2x speedup while full motion compensation gives marginally better results. These speedup factors are multiplied to the speedups provided by irradiance caching, a technique far faster than straight Monte Carlo pathtracing. Image frames were computed using an ambient accuracy setting of 15% and an ambient sampling density of 2048 samples per irradiance value at a resolution of 512x512. For comparison purposes, a reference solution and a perceptually accelerated solution are rendered at a higher resolution (640x480) and a sampling density of 8192 samples per irradiance value (Aleph Map Hi Res). As seen on the graph (Aleph Map vs. Aleph Map Hi Res), the acceleration is largely independent of the number of samples shot, because the perceptual solution changes only the spacing of the samples but not the sampling density. The reference solution takes between four to six hours to compute.The Aleph Map solution takes between 20 minutes to an hour to compute. Times are for a quad processor 550 MHZ Pentium III node.

512x512, with 2048 samples per irradiance value. Aleph Map Hi-res uses the Saliency map for velocity compensation, but rendered the scene at 640x480, 8192 samples per irradiance value. In most of the frames we achieve a 6x to 11x speedup over standard irradiance caching. Using spatial factors only we achieve a 2x speedup. A marginal improvement over spatial sensitivity is obtained if the full motion compensation heuristic is used in conjunction with spatiotemporal sensitivity. Note that all these improvements are compared to the speed of the unaugmented irradiance caching technique, which is hundreds of times more efficient than simple path tracing techniques. In addition, the speedup was found to be largely independent of the number of samples shot.

In this demonstration, we maintained good sampling protocols. The sampling density for each irradiance value is left unchanged, but the irradiance cache usage is perceptually optimized. Figure 6.11 shows the locations in the image at which irradiance values were actually computed. Bright spots indicate that an irradiance value was calculated while dark regions are places where the cache was used to obtain an interpolated irradiance value. This also explains why the speedup is independent of the number of samples shot, because the spacing of the irradiance cache is optimized, not the number of samples per irradiance value.

In static scenes where only the camera moves, the irradiance cache can be maintained over consecutive frames. Our technique was found to perform well even when such interframe coherence is used. An Antique Room sequence was rendered at a resolution of 640x480, 512 samples for the direct lighting and 8192 samples for the irradiance value (indirect lighting). The high number of direct samples is due to the presence of glossy surfaces in the scene. In the Pool and Art Gallery sequences, the surfaces were either specu-

Reference
Sampling
Density

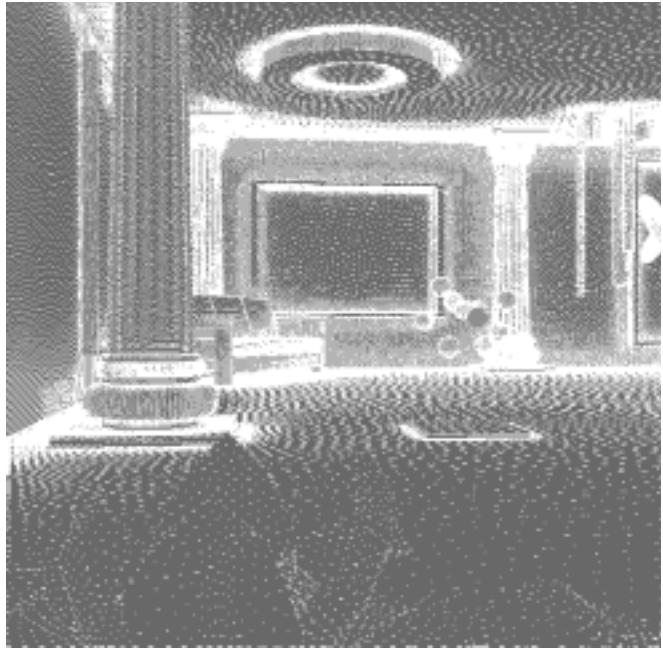

Aleph Map
Assisted
Sampling
Density

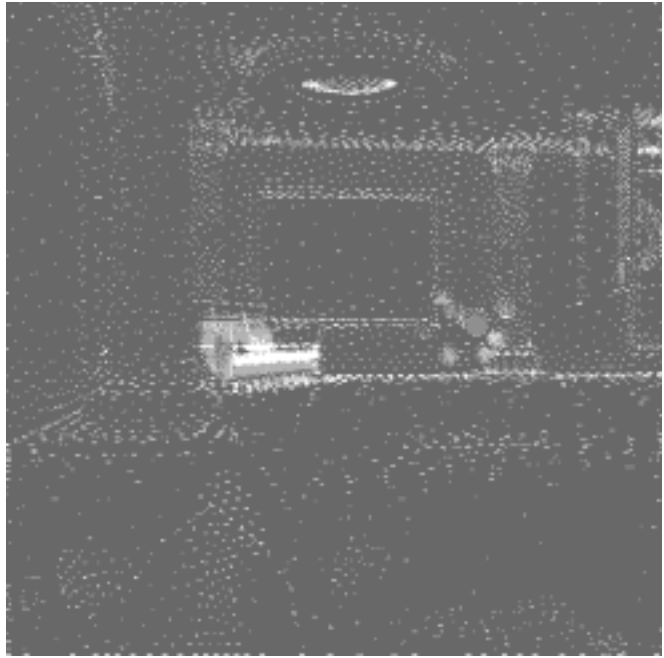

Figure 6.11: Sampling patterns for frame 0 of the Art Gallery sequence

The bright spots indicate where the irradiance value for the irradiance cache is generated and the dark spots indicate where an interpolated irradiance value is used. More irradiance values are needed near object boundaries and highly curved surfaces as it is at those locations that the error term is large.

lar or diffuse, obviating the need for large numbers of direct lighting samples. Figure 6.12 shows a frame from the Antique Room sequence.

Glossy surfaces require a large number of samples to evaluate the lighting solution correctly. The irradiance cache cannot be used for glossy surfaces because unlike Lambertian, diffuse surfaces, glossy surfaces require the evaluation of the glossy Bi-directional Reflectance Distribution Function for each incoming light value. Therefore, in the Antique Room sequence, only the diffuse term of the lighting solution is accelerated via the Aleph Map enhanced irradiance cache. The speedup of about six is less than previous scenes mainly due to the presence of glossy surfaces. The next section addresses the issue of applying the Aleph Map to progressive global illumination and will address the issue of accelerating the rendering of scenes with glossy surfaces.

Reference Solution

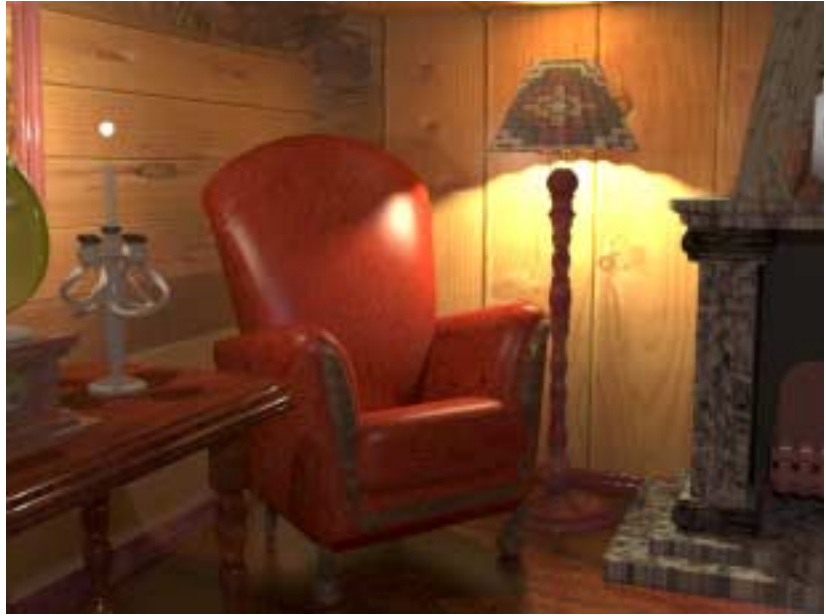

Aleph Map Accelerated Solution

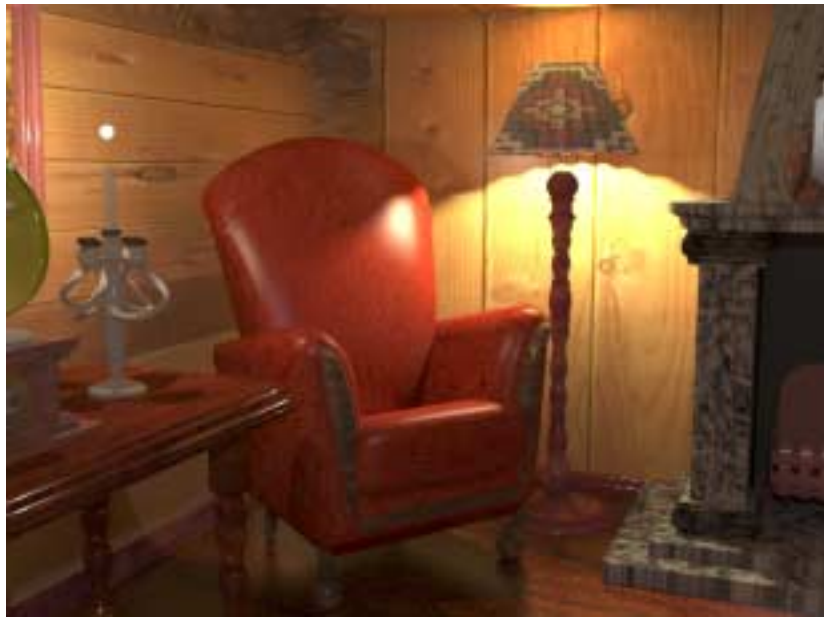

Figure 6.12: Antique Room Visual Comparison, Frame 32

The reference solution (top) and the Aleph Map accelerated solution (bottom) are presented for visual comparison. The Aleph Map accelerated solution was rendered six times faster than the reference solution.

## 6.2 Application to Progressive Global Illumination

In perceptually-driven progressive global illumination, programs utilize a variety of perceptual convergence tests to determine the stopping condition of a lighting solution. These perceptual convergence tests combine a statistical test method with a perceptually-based modification. The previous work section covers a whole array of these convergence tests. We will focus on the application of the Aleph Map in convergence testing.

Recall that the Aleph Map represents spatiotemporal sensitivity. That means, we can derive the physical convergence criteria simply by calculating the luminance threshold $\Delta L$ as follows:

$$\Delta L(x, y) = \aleph(x, y) \times \Delta L_{TVI}(L(x, y)) \qquad (6.9)$$

where $\Delta L(x,y)$ is the luminance threshold, $L(x,y)$ is the adaptation luminance calculated as the average luminance in a one degree diameter solid angle centered around the fixating pixel and $\Delta L_{TVI}$ is the threshold vs. intensity (TVI) function defined in Ward-Larson et al. [Ward97]. The TVI function tells us for a given adaptation luminance, the smallest contrast needed to discern one pattern from another. Figure 6.13 shows the shape of the TVI function.

In the style of Ramasubramanian et al. [Rama99], we can use the Aleph map to determine the stopping condition of a lighting solution by comparing the difference between two consecutive stages, N and N+1, of the lighting solution. The stopping condition is given in the following equation:

$$\left| L_N(x, y) - L_{N+1}(x, y) \right| < \Delta L(x, y) \qquad (6.10)$$

where $L_N(x,y)$, $L_{N+1}(x,y)$ are the mean pixel luminances from stages N and N+1 respectively of the lighting solution, and $\Delta L(x,y)$ is the luminance

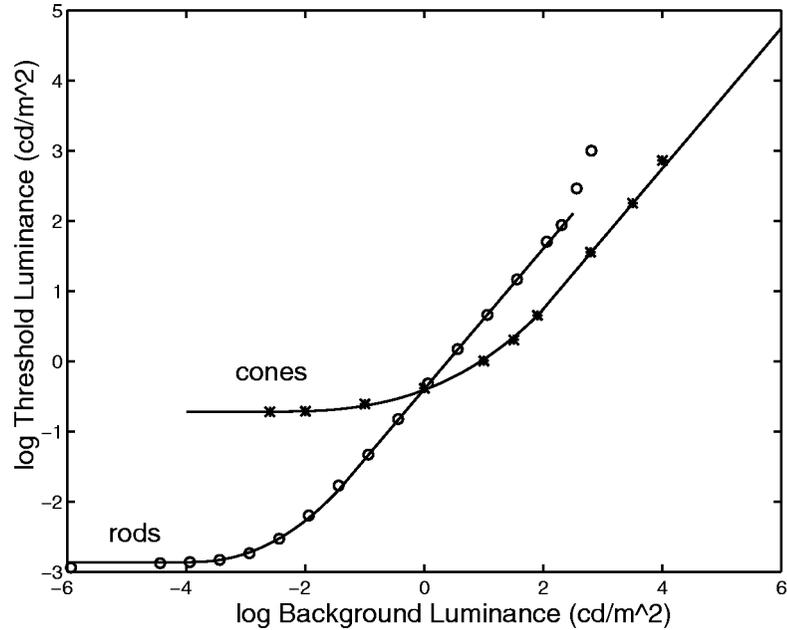

Figure 6.13: Threshold vs. Intensity (TVI) Function

Adapted from [Ferw96]. The TVI is a psychophysically derived function tells us the threshold luminance needed to detect a target from a background of a specified luminance. The Ward function is the envelope of the TVI curves for the cones and rods of the human eye.

threshold calculated in Equation 6.9. The rendering stops when the difference between two stages of the lighting solution is below the luminance threshold specified by the Aleph map and the TVI function.

Another way to use the luminance threshold presents itself in the calculation of the variance of a sample. Bolin and Meyer used the upper and lower bounds images derived from the variance of the lighting solution as boundary images in their paper [Boli98]. Lee et al. [Lee85] proposed a stopping condition based on variance which we will extend to the perceptual domain using the Aleph Map. Recall that the sample variance of a set of N identically distributed samples of a random variable, $V_N$, may be calculated as:

$$V_N = \frac{1}{N}\left(\sum_{i=1}^{N} L_i^2 - \frac{1}{N}\left(\sum_{i=1}^{N} L_i\right)^2\right) \qquad (6.11)$$

where $L_i$ is the pixel luminance at iteration i of the lighting solution. The simplest stopping condition is to ensure that the standard deviation of the pixel luminance is smaller than the luminance threshold:

$$\sqrt{V_N(x, y)} < \Delta L(x, y) \qquad (6.12)$$

where $\Delta L(x,y)$ is the luminance threshold calculated in Equation 6.9. We shall call this test the "Aleph Variance Test (AVT)".

Kirk and Arvo [Kirk91] point out that having a stopping condition based on the variance of a sample can introduce a systemic bias to a lighting solution. They suggest a simple rule to avoiding the bias by determining how a sample is to be used before it is drawn rather than basing the decision on the actual samples being drawn. In the light of this observation, we may modify our sampling protocol in the following manner:

$$\text{No. samples shot} = \frac{\text{Maximum number of samples}}{\aleph(x, y)} \qquad (6.13)$$

The sampling protocol described in Equation 6.13 is a simple heuristic that we shall call the "Aleph Sampling Protocol (ASP)".

We re-render the Antique Room with the Aleph Variance Test and the Aleph Sampling Protocol in order to determine the effectiveness of these two techniques. The Antique Room contains many glossy surfaces, requiring that many samples be shot per pixel in order to get the correct integration over the glossy reflectance function. The reference solution shoots 512 rays per pixel at all locations, and 8192 samples per irradiance value for indirect illumination. The Aleph Variance Test (Equation 6.12) solution shoots as many rays as it needs until the standard deviation (square root of variance) is below the luminance threshold or the limit of 512 is reached. The Aleph Sampling Protocol (Equation 6.13) shoots (512 / Aleph Map value) rays or 16 rays, whichever is bigger. Figure 6.14 shows the timings associated with using these two speedup techniques in comparison with the unaugmented irradiance cache and the Aleph Map enhanced irradiance cache in rendering of the Antique Room. In the top chart, the timings are for the initial frame where the irradiance cache is filled. In this phase of the rendering, the filling of the irradiance cache takes up most of the computation time, which is why the Aleph Map enhanced Irradiace cache, the Aleph Variance Test and the Aleph Sampling Protocol perform at about the same speed. Since this is a static scene, the irradiance cache can be carried over from the previous frame, unlike the earlier scenes where there were moving objects. When the irradiance cache is already filled, indirect diffuse illumination computation is efficient and the majority of the computation is required to solve for the glossy direct reflections. In this scenario, as shown in the bottom chart of Figure 6.14, the Aleph Variance Test and the Aleph Sampling Protocol are more efficient than the Aleph Map enhanced Irradiance Caching because they accelerate the rendering of glossy direct illumination as well. The two techniques also display an order of magnitude speedup in comparison with unaugmented irradiance caching. Figure 6.15 shows a images of the Antique Room rendered with the Aleph Variance Test and the Aleph Sampling Protocol.

One interesting observation is that the error in the image has been pushed into visually non-salient locations. The errors are not noticeable from the proper viewing distance, but a magnification of a factor of eight reveals the locations to which the errors have been displaced. Figure 6.16 shows a magnification of the glossy region of the floor under the armchair in the Antique Room. These errors are not visible at full resolution, but become apparent

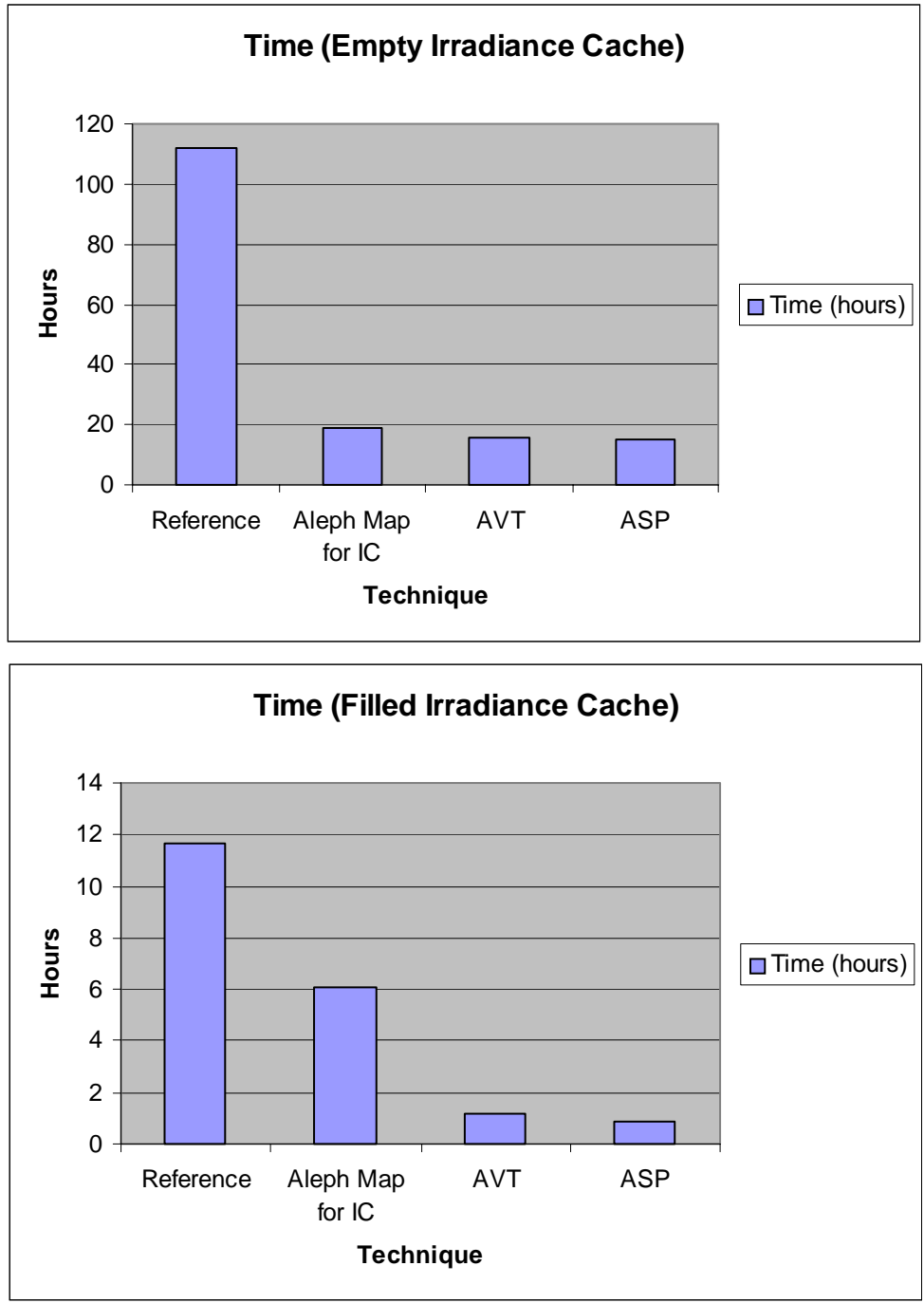

Figure 6.14: Antique Room Lighting Speedups

The top chart compares the timings per frame when the irradiance cache is empty. Since this scene is static, the irradiance cache can be recycled through subsequent frames. In the bottom chart, the AVT and ASP pull ahead because they perform perceptually-based progressive global illumination.

Aleph Variance Test (AVT)

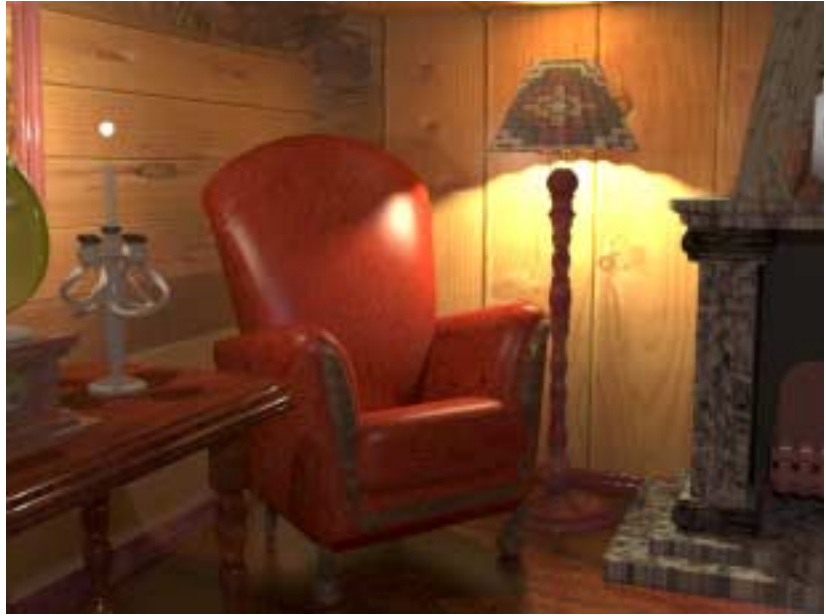

Aleph Sampling Protocol (ASP)

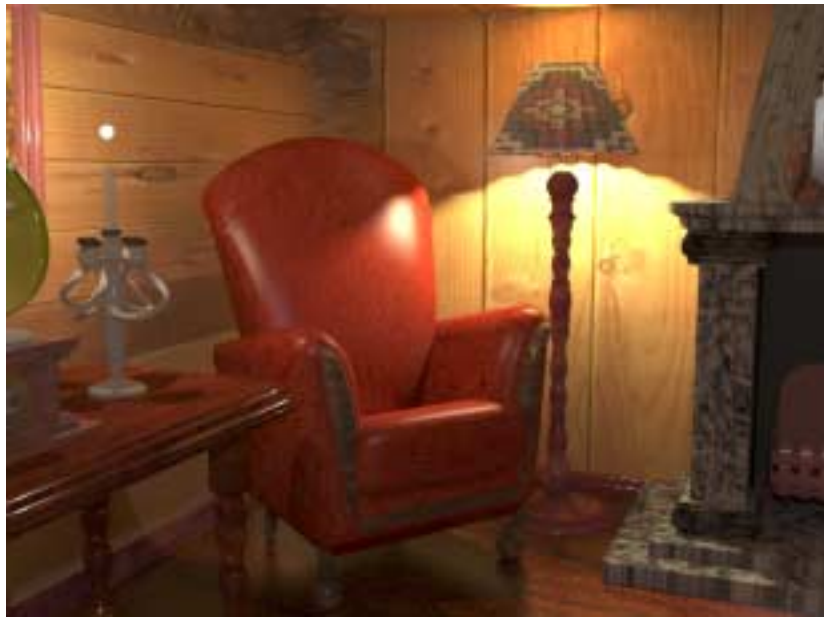

Figure 6.15: Antique Room Progressive Rendering

The Aleph Variance Test (top) and the Aleph Sampling Protocol (bottom) are presented for visual comparison against the non-progressive rendering techniques in Figure 6.12 (page 71).

Reference
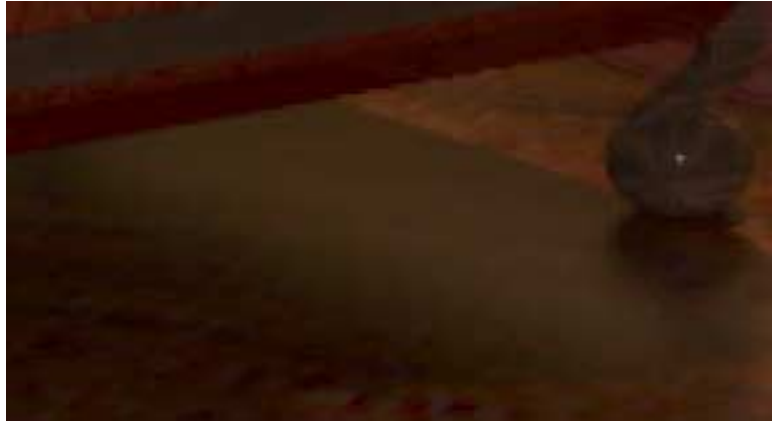

Aleph
Variance
Test
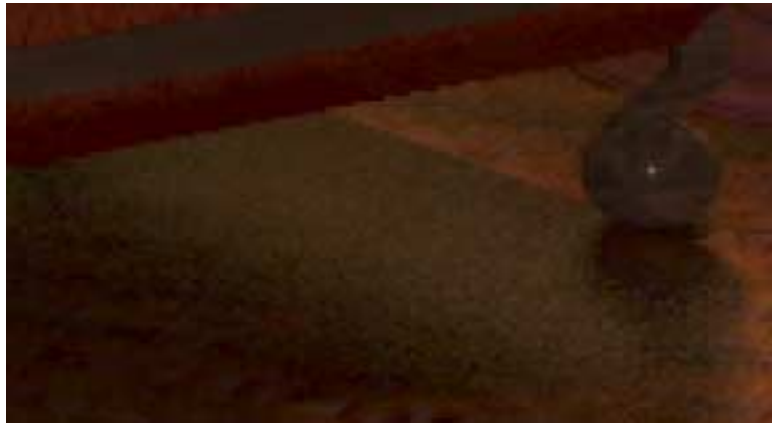

Aleph
Sampling
Protocol
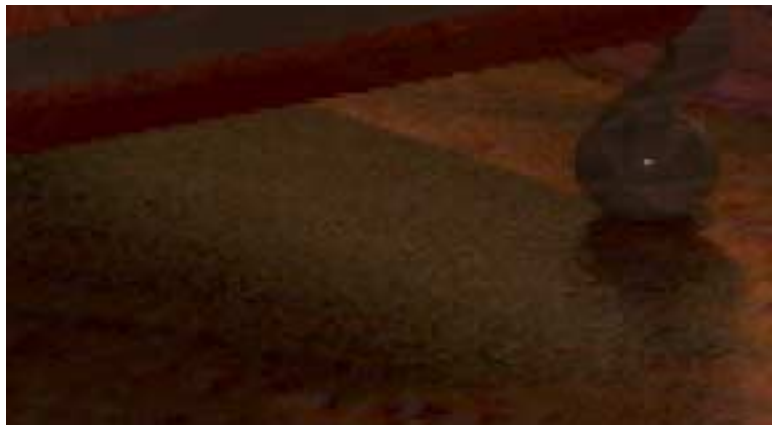

Figure 6.16: Antique Room Magnification

This figure shows a region of the floor of the antique room underneath the armchair, magnified by a factor of eight. The reference solution shows a smooth glossy reflection off the floor whereas the AVT and the ASP solutions have noticeable noise. The noise is below perceptual threshold at normal magnifications and is also 'pushed' to less important locations in the image.

upon magnification. The Aleph Map enhanced Irradiance Cache does not share this property as it samples direct illumination with the same amount of samples as the unaugmented irradiance cache.

# CHAPTER 7
# Discussion

*The words of the Preacher, the son of David, king in Jerusalem: "...of making many books there is no end; and much study is a weariness of the flesh. Let us hear the conclusion of the whole matter: Fear God, and keep his commandments: for this is the whole duty of man. For God shall bring every work into judgment, with every secret thing, whether it be good, or whether it be evil"*

## 7.1 Validation

In order to test the efficacy of the Aleph Map independent of any acceleration technique, we compute the luminance threshold as in Equation 6.9 and multiply it with a uniform random number distributed over (0.0 .. 1.0) in order to derive a sub-threshold noise map. If the theory holds, the addition of the sub-threshold noise map should not be discernible to the viewer. Figure 7.1 shows how a noisy reference solution is created for the art gallery sequence. A video of the reference solution and the noisy reference solution were shown to a panel of viewers and it was found that the noise was indeed sub-threshold in normal viewing conditions. This is an empirical test of the Aleph map and is not meant to be a rigorous psychophysical examination of the map, which the author is not qualified to conduct. The Saliency Map was validated separately by Itti and Koch [Itti00], and the Spatiotemporal Sensitivity was validated by Daly [Daly98].

Reference
Solution

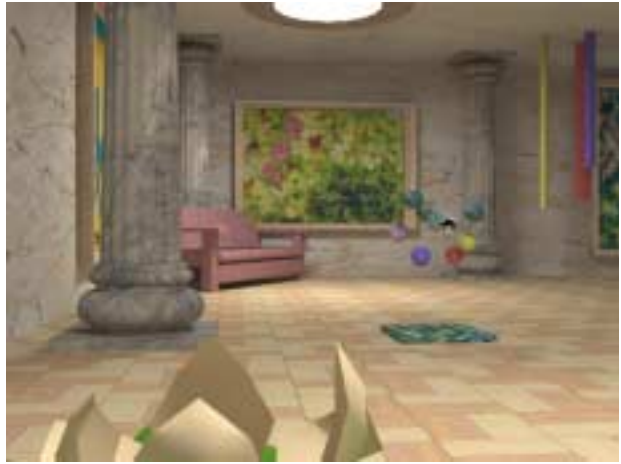

+ Aleph Map
* TVI
* Unit Random Noise

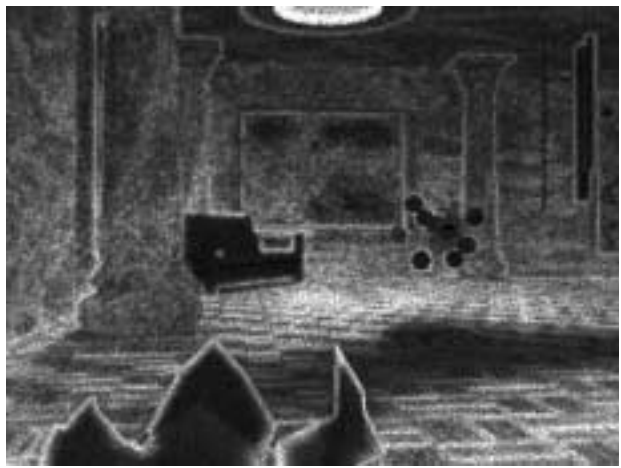

= Noisy
Reference

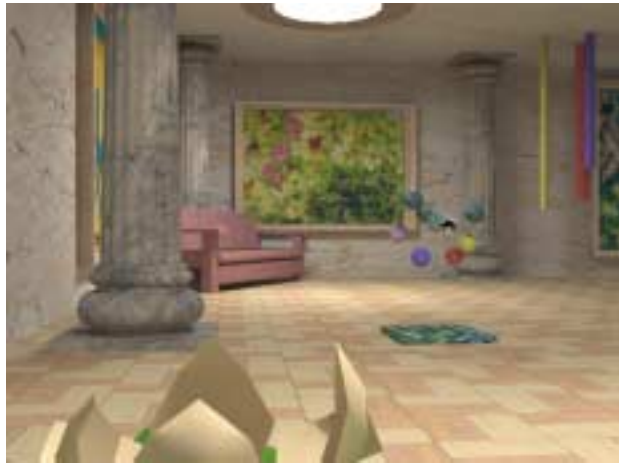

Figure 7.1: Noise Map Test

A noise map is constructed using the Aleph Map, the TVI and a unit random
number. This noise map created for each frame of the reference solution and
added to the reference frame to create a 'noisy' reference solution.

## 7.2 Discussion

In viewing the Art Gallery sequence, it was discovered that repeated viewings can cause the viewer to pay more attention to unimportant regions. In doing so, the viewer deliberately chose to ignore attention cues and focus on unimportant areas such as the ceiling. This introduces a top-down behavioral component to visual attention that is not accounted for in our model. The pool table sequence had unambiguous salient features (the pool balls) and was not as susceptible to the replay effect. One assumption we made was that the rendering technique does not introduce visually salient errors, which might not hold true for some kinds of rendering techniques.

Visual sensitivity falls rapidly as a function of foveal eccentricity. An experiment incorporating foveal eccentricity into the model was performed, and significant speedup was achieved. However, the animations generated with the use of foveal eccentricity tended to be useful only in the first few runs of the animation, as viewers tended to look away from expected foveal regions once they had seen the animation a number of times. Visual artifacts are visible once non-foveal regions are subject to the close scrutiny of observer.

An assumption that we make is that there is something in the scene to draw an observer's attention to. This implies that our technique would not work as well when there is nothing in a scene that stands out or when everything stands out equally. When such cases are detected, we would suggest defaulting to motion compensating the entire scene and sacrifice performance for accuracy.

The current implementation of the perceptual metric has complete knowledge of the animation to be rendered in the form of image estimates. In inter-active applications, such information is not readily available. In such cases we suggest using motion prediction for estimating image plane velocity and moving blocks of sprites using the motion prediction to derive the image estimate for the next frame. Additionally, the processing overhead, though negligible for global illumination, would pose a problem for real-time applications. One possible solution is to use the graphics hardware as a SIMD engine for computing the Aleph Map. This would require the image convolution extension, the color matrix extension and extended range color channels, features that are soon to appear on commodity graphics hardware.

Our implementation also does not include color and orientation in the sensitivity computation, although those factors are considered in the computational model of visual attention. We also do not implement contrast masking. This makes our model conservative, but it is better to err on the safe side. We have chosen to treat each component of the visual system as multiplicative with each other and the results have shown that it works but the human visual system is non-linear and has vagaries that would be hard to model.

## 7.3 Future Work & Conclusion

There are many areas that the Aleph Map may be applied to that have not been explored in this thesis. One such area would be video compression. The Aleph Map can be used to perceptually guide the quantization of image frames in a sequence prior to video compression. Another application would be to select the Level of Detail in real time rendering applications. For example, a heuristic could be constructed such that when the Aleph map has small values, a more detailed level should be used for that area of the scene. It would be interesting to apply the technique to view-independent global illumination as well.

In this thesis, a novel perceptual technique for exploiting the limitations of the Human Visual System with regards to spatiotemporal sensitivity is shown. Previous perceptual techniques functioned well for static scenes, whereas our technique breaks new ground by enhancing perceptual techniques and applying them to dynamic environments. The new technique takes the form of a spatiotemporal error threshold elevation map that is modified by a computational model of visual attention and adapted to specific applications. The resulting Aleph Map can be used as a perceptual oracle to guide global illumination via optimizing irradiance caching. Its use was demonstrated in progressive illumination as well as a perceptual oracle for glossy direct illumination. The algorithms enhanced by our perceptual model exhibited an order of magnitude increase in efficiency. The possibilities for Aleph Map use are many and varied. The author is pleased to make this contribution to the state of the art in graphics research and technology.